\documentclass[journal=jacsat,manuscript=article]{achemso}

\usepackage{chemformula} % Formula subscripts using \ch{}
\usepackage[version = 4]{mhchem} % Formulas using \ce{}
\usepackage[T1]{fontenc} % Use modern font encodings
\usepackage{threeparttable, tablefootnote}

\usepackage{hyperref}

\usepackage{graphicx}
\usepackage{float}
\newfloat{scheme}{htbp}{los}
\floatname{scheme}{Scheme}
\floatname{chart}{Chart}
\newfloat{graph}{htbp}{loh}
\usepackage[dvipsnames]{xcolor}
\usepackage{amsmath}
\usepackage{amstext}
\usepackage{amssymb}

\author{Marissa Vlasblom}
\affiliation[Leiden]
{Leiden Observatory, Leiden University, 2300 RA Leiden, Netherlands}
\email{vlasblom@strw.leidenuniv.nl}

\author{Aditya M. Arabhavi}
\affiliation[Groningen]{Kapteyn Astronomical Institute, Rijksuniversiteit Groningen, Postbus 800, 9700AV Groningen, Netherlands}

\author{Niels de Klerk}
\affiliation[Leiden]
{Leiden Observatory, Leiden University, 2300 RA Leiden, Netherlands}

\author{Inga Kamp}
\affiliation[Groningen]{Kapteyn Astronomical Institute, Rijksuniversiteit Groningen, Postbus 800, 9700AV Groningen, Netherlands}

\author{Beno\^it Tabone}
\affiliation[Paris]{Universit\'e Paris-Saclay, Universit\'e Paris Cit\'e, CEA, CNRS, AIM, F-91191 Gif-sur-Yvette, France}

\author{Ewine F. van Dishoeck}
\affiliation[Leiden]
{Leiden Observatory, Leiden University, 2300 RA Leiden, Netherlands}
\alsoaffiliation[MPE]{Max-Planck Institut f\"ur Extraterrestrische Physik (MPE), Giessenbachstr. 1, 85748, Garching, Germany}

\title{Detecting nitrogen-carriers in the inner regions of protoplanetary disks}

\begin{document}

% \maketitle

\begin{abstract}
  Nitrogen is a key element for building habitable worlds, yet only a small fraction of the available N-budget of planet-forming disks has been detected. In particular, the lack of any IR \ce{NH3} detection is striking, as this molecule is predicted to be rather abundant in the warm, inner regions of protoplanetary disks, and therefore potentially readily incorporated into (giant) planets' atmospheres. We present a combined modeling and observational study of N-bearing molecules in planet-forming disks, using detailed thermo-chemical disk models that investigate the sensitivity of N-containing molecules to the bulk elemental composition of the disk. Our models predict a strong increase in HCN flux with high C/H, and conversely a strong increase in flux from NO when O/H is high. The flux from \ce{NH3} is not very sensitive to O/H, but does decrease at high C/H due to competition with HCN. However, the absolute \ce{NH3} flux predicted by our model is not large enough to be detected with JWST-MIRI, even when N/H is enhanced by an order of magnitude. The flux from NO, on the other hand, is potentially detectable, and could therefore provide further insights into the N-budget of the inner disk. Using a cross-correlation technique, we search for \ce{NH3} and NO detections in three disks, GW Lup, Sz 98, and V1094 Sco. We do not find any \ce{NH3} detections, and only one tentative NO detection in V1094 Sco, though this needs further study to be confirmed. Additionally, we demonstrate that future facilities in the FIR may provide a better opportunity to detect \ce{NH3} and thereby draw a comparison to the \ce{NH3} budget known to be present in interstellar ices.
\end{abstract}

\section*{Keywords}
Astrochemistry - IR spectroscopy - Protoplanetary disks - Nitrogen chemistry - Ammonia \ce{NH3} - Modeling
% All Articles, Letters, Reviews, and Perspectives should include 5-8 keywords

\section{Introduction}

Planets form in rotating disks of gas, dust and ice around young stars, just as the planets in our own Solar System were built some 4.5 billion years ago from the pre-solar nebula. Thus, it is the chemical make-up of disks that sets the overall elemental and molecular composition of new planets \citep{Oberg21,vanDishoeck06}. The chemistry in disks has been an active topic of study over the past 30 years (see \citet{Oberg23} for review), with Eric Herbst and his collaborators having a pioneering role in its development \citep{Aikawa99a,Aikawa02,Herbst14}.

With the advent of the Atacama Large Millimeter/sub-millimeter Array (ALMA), the chemistry in the cold ($<$100 K) outer disk has become accessible to detailed observational studies ($>$20 au\footnote{1 au = 1.5$\times 10^{13}$ cm = Sun-Earth distance}). Simple and more complex C, O, N and S-containing molecules have been detected through their rotational transitions \citep{McGuire22,Booth24}. Their chemistry is thought to be controlled by the balance between ion-molecule reactions, photodissociation, freeze-out and grain surface formation (ice chemistry) followed by thermal and non-thermal desorption of ices \citep{Aikawa02,Walsh14,Oberg23, henning2013}. However, the majority of the detected exoplanet population are formed in the inner 10 au of the disk \citep{Dawson18,Drazkowska23} that is inaccessible to ALMA. This region is characterized by higher temperatures and densities, triggering a different chemistry \citep{walsh2015, woitke2018}.

Thanks to the {\it James Webb Space Telescope} (JWST\cite{rigby2023}), the chemistry in this warm inner disk can now be studied in detail \citep{kamp2023, vandishoeck2023, banzatti2023, pontoppidan2024, arulanantham2025}, greatly improving on pioneering observations with the {\it Spitzer Space Telescope} (see \citet{Pontoppidan14} for review). The JWST mid-infrared spectrometer covers the 5--28 $\mu$m wavelength range at a spectral resolving power $R=\lambda/\Delta \lambda$ {of $\sim$3500 at the shortest wavelengths to $\sim$1500 at the longest wavelengths. This }is sufficient to separate individual ro-vibrational lines and detect weak features on top of a strong continuum \citep{Wright23}. Infrared spectroscopy has the advantage that molecules without a permanent dipole moment, such as CH$_4$, C$_2$H$_2$ and CO$_2$, can be observed in addition to other simple molecules that also have strong infrared features, most notably H$_2$O. Indeed, JWST spectra of planet-forming disks show a wealth of lines, ranging from disks rich in water \citep{temmink2025, banzatti2025, arulanantham2025} to those rich in hydrocarbon molecules including even benzene \citep{tabone2023,arabhavi2024,Arabhavi25a,Long25, kanwar2024_sz28, colmenares2024, perotti2026}.

Most chemical studies of disks to date have focused on the carbon and oxygen-containing molecules. However, nitrogen is also a key element for building a habitable planet as a fundamental component of nucleic acids and proteins \citep{Krijt23}. HCN and HC$_3$N are readily detected in infrared spectra of disks \citep{arulanantham2025,Arabhavi25a, salyk2025}, and a handful of N-containing molecules are observed in the outer disk \citep{Bergner21,Oberg23}, including cold \ce{NH3} \citep{salinas2016}. However, these molecules contain only a small fraction (few \%) of the available N-budget. The bulk of the gaseous nitrogen is thought to be contained in N and N$_2$ \citep{Schwarz14,walsh2015, kanwar2025}, two species that are alas invisible at both infrared and millimeter wavelengths. Little is known about other N-containing molecules that could be abundant in the inner disk, most notably warm NH$_3$ \citep{walsh2015}. {So far,} gaseous \ce{NH3} has {only} been detected in absorption at mid-infrared wavelengths in a protostellar hot core with JWST \citep{vangelder2024} (with an abundance that is consistent with ice sublimation) and in the surface layers of a Class I disk \citep{najita2021} (with an abundance similar to that of HCN). \citet{kaeufer2024} present a tentative \ce{NH3} detection in the hydrocarbon-rich very-low-mass star (VLMS) disk Sz 28. \citet{pontoppidan2019} however searched for \ce{NH3} in three T Tauri disks with high spectral resolution, and were able to provide only stringent upper limits on the \ce{NH3} abundance, predicting it to be lower than typical ice abundances.

We here present a combined modeling and observational study of NH$_3$ and other N-bearing molecules in the warm inner regions of planet-forming disks around solar-mass stars in order to probe the nitrogen chemistry in disks. We use a detailed thermo-chemical disk model to investigate the sensitivity of the abundances of N-bearing species to a range of elemental abundances of C, O and N, as was recently done to investigate the sensitivity of the H$_2$O, CO$_2$, and C$_2$H$_2$ chemistry \citep{arabhavi2026}. These abundances are varied to mimic the effects of drifting icy grains from the outer disk, with ices sublimating at their so-called ice lines and thereby enhancing the inner disk gas with elements that are important for chemistry \citep{kalyaan2021, krijt2025}. However, icy pebbles can also be trapped in dust rings outside their ice lines \citep{Pinilla12} that are ubiquitously observed with ALMA in the outer disk \citep{Andrews20}. This would lead to a decrease in heavy elements (C, O, N) in the inner disk, as most of the original gas is rapidly advected onto
the star \citep{Booth19}. In addition to the chemistry, the corresponding mid-infrared spectra can also be simulated from the models for direct comparison with observations.

In this work, we use data from the JWST Mid-INfrared Disk Survey (MINDS) program \citep{henning2024} to perform a deep search for NH$_3$ and chemically related molecules like NO. NH$_3$ is a particularly interesting molecule since it is a well-known component of interstellar ices with an abundance of $\sim$6\% with respect to H$_2$O ice \citep{Bottinelli10} and twice that amount contained in NH$_4^+$ in salts \citep{boogert2015}. These ices are expected to be transported largely unaltered from the cold collapsing cloud to the outer disk where they are seen in comets \citep{Altwegg19,Altwegg22}. When drifting icy pebbles cross the ammonia ice line, close to that of water, they release NH$_3$ back in the gas which could lead to enhanced N abundances. This sublimated gas may then be vertically mixed upwards into the upper layers where chemical timescales are typically very short, and therefore the librated \ce{NH3} may also be converted into other N-bearing species. Evidence for ice sublimation at the ice line has been claimed for H$_2$O \citep{banzatti2023} and is also predicted for the case of CO$_2$ \citep{bosman2017, vlasblom2025_CXTau, sellek2024}. 

Since the bulk of the gaseous nitrogen budget is contained in \ce{N2} (and N in the surface layers), the strong \ce{N2} bond needs to be broken to liberate the atomic nitrogen required for building other N-containing molecules like NH$_3$ and NO. This can be achieved by reactions with He$^+$, which in turn is produced by X-rays or highly-energetic particles like cosmic rays \citep{Agundez08,walsh2015}. Thus, constraints on the NH$_3$ content of disks, in combination with those of NO and HCN, could provide insight into various chemical and physical processes.

The outline of this paper is as follows. In \S 2, we describe our modeling setup. We present the main results regarding the abundances, fluxes, and observability of the main N-carriers, along with possible future prospects for detecting \ce{NH3} in the FIR in \S 3. In \S 4 we discuss their implications, and \S 5 summarizes our main conclusions. 

\section{Methods}

{This work uses} \underline{Pro}toplanetary \underline{Di}sk \underline{Mo}del (P{\small RO}D{\small I}M{\small O} \citep{woitke2016,Kamp17, woitke2009}) to simulate a standard T Tauri disk with different elemental abundances and the Fast Line Tracer (FLiTs \citep{woitke2018}) to simulate the mid-infrared spectra from the disk models. These models calculate the radiation field and the dust temperatures in the disk consistently, followed by the chemistry and energy balance of the gas. 

{We create two grids of models, following the same approach as the IDLI (Infrared Diagnostics from Line emIssion) grid presented in \citet{arabhavi2026}. The reference model is the fiducial T Tauri model from \citet{woitke2016, woitke2018}, consisting} of a disk of mass 0.02\,$M_{\odot}$ around a 0.7\,$M_{\odot}$ star with one solar luminosity{, which is fairly representative of a typical T Tauri disk\citep{manara2023}}. {Contrary to \citet{arabhavi2026}, the total disk gas-to-dust ratio used in all models throughout this work is 1000 rather than 100}, as this has been found to better reproduce observed line fluxes \citep{woitke2024, meijerink2009}. {Dust settling is accounted for following the description in \citet{riols_lesur_2018} assuming a turbulent mixing parameter of $\alpha=10^{-3}$.} The elemental composition of {this} reference model is taken from Kamp et al. \citep{Kamp17}, which yields a C/O ratio of 0.45. {Then,} the carbon and oxygen elemental abundances are each varied by -0.5, -0.25, 0, 0.25, and 0.5 dex (corresponding to a factor 0.3, 0.6, 1., 1.8, 3.2), resulting in a total of 25 models {that make up the `solar-N/H' grid. These variations in C/H and O/H represent enhanced abundances in the inner disk caused by drifting icy pebbles or decreased abundances due to dust traps beyond snow lines.} As the emission from nitrogen-bearing species (other than HCN) is generally weaker than that of dominant oxygen- and carbon-bearing species (such as $\rm H_2O$, CO, $\rm CO_2$), we re-run the same grid of models as outlined above, where we now enhance the elemental abundance of nitrogen by an order of magnitude (1 dex). {This grid is referred to as the `enhanced-N/H' grid. The elemental composition of the two reference models of the two grids (the two models with the fiducial, solar C/H and O/H values) are presented in Table \ref{tab:comp_ref}. The changes in elemental composition are in turn also} able to affect the thermal balance of the gas, and therefore the temperature structure and emission spectra of the disk. The full list of abundances, along with more details about the disk properties, can be found in \citet{arabhavi2026}.

% The grid of models are taken from \citet{arabhavi2026}. However, in this paper we assume a total disk gas-to-dust ratio of 1000 contrary to 100 in \citet{arabhavi2026} as this has been found to better reproduce observed line fluxes \citep{woitke2024, meijerink2009}.  The disk structure remained the same across the grid, including the dust properties. The original grid presented in \citet{arabhavi2026} varies in carbon and oxygen elemental abundances, which in turn are able to affect the thermal balance of the gas and therefore the temperature structure and emission spectra of the disk. 

% The reference model consists of a disk of mass 0.02\,$M_{\odot}$ around a 0.7\,$M_{\odot}$ star with one solar luminosity\citep{woitke2016,woitke2018,woitke2024}. The elemental composition of the reference model is taken from Kamp et al. \citep{Kamp17}, which yields a C/O ratio of 0.45. We highlight the most important values in Table \ref{tab:comp_ref}. The full list of abundances, along with more details about the disk properties, can be found in \citet{arabhavi2026}. The carbon and oxygen elemental abundances are each varied by -0.5, -0.25, 0, 0.25, and 0.5 dex (corresponding to a factor 0.3, 0.6, 1., 1.8, 3.2), resulting in a total of 25 models. As the emission from nitrogen-bearing species (other than HCN) is generally weaker than that of dominant oxygen- and carbon-bearing species (such as $\rm H_2O$, CO, $\rm CO_2$), we re-run the same grid of models as outlined above, where we now enhance the elemental abundance of nitrogen by an order of magnitude (1 dex).

\begin{table}
  \caption{Elemental composition of the reference {models of the two grids}}
  \label{tab:comp_ref}
  \centering
  \begin{tabular}{l c c}
    \hline
    \hline
     & {Solar N/H} & {Enhanced N/H} \\
    \hline
    Element & $n$(X)/$n_{\rm H}$ & $n$(X)/$n_{\rm H}$ \\
    \hline
    H & $1.0 \times 10^{0}$ & $1.0 \times 10^{0}$ \\
    He & $9.64 \times 10^{-2}$ & $9.64 \times 10^{-2}$ \\
    C & $1.38 \times 10^{-4}$ & $1.38 \times 10^{-4}$ \\
    O & $3.02 \times 10^{-4}$ & $3.02 \times 10^{-4}$ \\
    N & $7.94 \times 10^{-5}$ & $7.94 \times 10^{-4}$ \\
    \hline
  \end{tabular}
\end{table}

{The chemical network used in this work, as well as in \citet{arabhavi2026}, is the large DIsc ANAlysis (DIANA) chemical network \citep{Kamp17} with the addition of the large hydrocarbon network\citep{kanwar2024_model}. The network contains 327 species and 4121 reactions, which are taken from the UMIST2012 database\citep{mcelroy2013_UMIST12}. The full details of the network can be found in \citet{Kamp17} and \citet{kanwar2024_model}. We solved the steady state chemistry in all models. }

Photodissociation of \ce{N2} is important for the nitrogen chemistry of disks as it allows atomic nitrogen to be liberated and potentially incorporated into other species. {In our models (as well as the models from \citet{arabhavi2026}),} the photodissociation cross sections considered come from \citet{heays2017}. The effects of \ce{N2} self-shielding {are accounted for }following \citet{li2013}. Additionally, reactions with vibrationally excited \ce{H2} are important for overcoming reaction barriers, in particular the barrier of the \ce{N + H2 -> NH + H} reaction at $12\,650$ K\citep{davidson1990}. Vibrationally excited \ce{H2}, denoted as H$_2^*$, is treated as a separate species in the chemical network, representing an excited state corresponding to a vibrational pseudo-level of $v=6$ with an energy of 2.6 eV. The FUV pumping of \ce{H2} into H$_2^*$ is assumed to proceed at a rate of 10 times the \ce{H2} photodissociation rate, and collisional excitation and de-excitation reactions of H$_2^*$ with H and \ce{H2} are implemented following \citet{woitke2009} and \citet{tielens1985}. {The stellar radiation field consists of a 4000 K blackbody, including UV excess due to accretion (see \citet{woitke2016}). We adopt an X-ray luminosity of $L_X=10^{30}$ erg s$^{-1}$ and a cosmic-ray ionization rate of $\xi_{\rm CR} = 1.7 \times 10^{-17}$ s$^{-1}$. }

The mid-infared spectra calculated from these models include more than 76000 lines, comprising rovibrational transitions of the infrared active bands of dominant nitrogen carriers such as $\rm NH_3$, $\rm NO$, and HCN. The molecular data for the nitrogen carriers are taken from the HITRAN database\citep{2022JQSRT.27707949G}. Of the species in our spectra, the emission from \ce{H2, H2O, O, CO}, and OH is treated in non-LTE, along with several atomic and molecular ions. A full list of the line transitions included in our models can be found in \citet{arabhavi2026}. The high resolution model spectra are convolved to the lower $R\approx3000$ spectral resolving power of MIRI for comparison with JWST data.

We also search for detections of N-bearing species in {three JWST-MIRI spectra of T Tauri disks (GW Lup\cite{grant2023}, Sz 98\cite{gasman2023b}, V1094 Sco; Tabone et al. subm.).} To this end, a cross-correlation technique is used.
For \ce{NH3}, both the $\nu_4$ bending mode at 6 $\mu$m (1659 cm$^{-1}$) and the $\nu_2$ umbrella mode at 10-12 $\mu$m (993 cm$^{-1}$) were considered, and for NO, the $\nu_1$ stretching mode at 5.3 $\mu$m (1886 cm$^{-1}$) is considered.
For each species, we take the predicted spectrum from our reference model as our template ($F_{\rm mod}$), resample it to the same wavelength grid as our data ($F_{\rm obs}$) and cross-correlate it with our data following
\begin{align}\label{eq:cc}
    CC({\rm lag}) = \int_{\lambda_1}^{\lambda_2} F_{\rm obs}(\lambda) F_{\rm mod}(\lambda + {\rm lag}) d\lambda
\end{align}
where the lag corresponds to the shift in wavelength between the data and the template spectrum, given in units of $\mu$m. This is compared to the auto-correlation of the template spectrum, which is defined as
\begin{align}\label{eq:ac}
    AC({\rm lag}) = \int_{\lambda_1}^{\lambda_2} F_{\rm mod}(\lambda) F_{\rm mod}(\lambda + {\rm lag}) d\lambda
\end{align}
and thus represents a pure signal. {To evaluate the significance of a cross-correlation signal, a procedure similar to a bootstrap method is used. To this end,} an array representing pure noise {($F_{\rm noise}$)} is created by randomly drawing samples from our data. We calculate the cross-correlation between our template spectrum $F_{\rm mod}$ and this randomized flux array $F_{\rm noise}$. {The significance of the signal is then estimated by calculating} the difference between the peak of the cross-correlation (at lag = 0 $\mu$m) and the median value in a small region around the peak {($\delta_{\rm noise}$)}. This step is repeated $10\,000$ times to obtain a distribution of these {$\delta_{\rm noise}$} values. Then, {every time} the cross-correlation between the data $F_{\rm obs}$ and the template spectrum {$F_{\rm mod}$} {is performed, the difference between the peak and the median, $\delta_{\rm obs}$, is calculated again.} The result is deemed significant (resulting in a detection of the molecule) if {$\delta_{\rm obs}$ is found to exceed 95\% of the $\delta_{\rm noise}$ values that are} generated with the randomized spectrum $F_{\rm noise}$.
% Additionally, an array representing pure noise is created by randomly drawing samples from our data. We calculate the cross-correlation between our template spectrum $F_{\rm mod}$ and this randomized flux array $F_{\rm noise}$. The difference between the peak of the cross-correlation (at lag = 0 $\mu$m) and the median value in a small region around the peak is estimated, to give a measure of the significance of the signal. This step is repeated $10\,000$ times to obtain a distribution of these values. Then, for the cross-correlation between the data $F_{\rm obs}$ and the template spectrum, the result is deemed significant (resulting in a detection of the molecule) if the difference between the peak and the median exceeds 95\% of the values generated with the randomized spectrum $F_{\rm noise}$.

\section{Results}

\subsection{Abundances and fluxes of N-carriers}

\begin{figure}
    \centering
    \includegraphics[width=\linewidth]{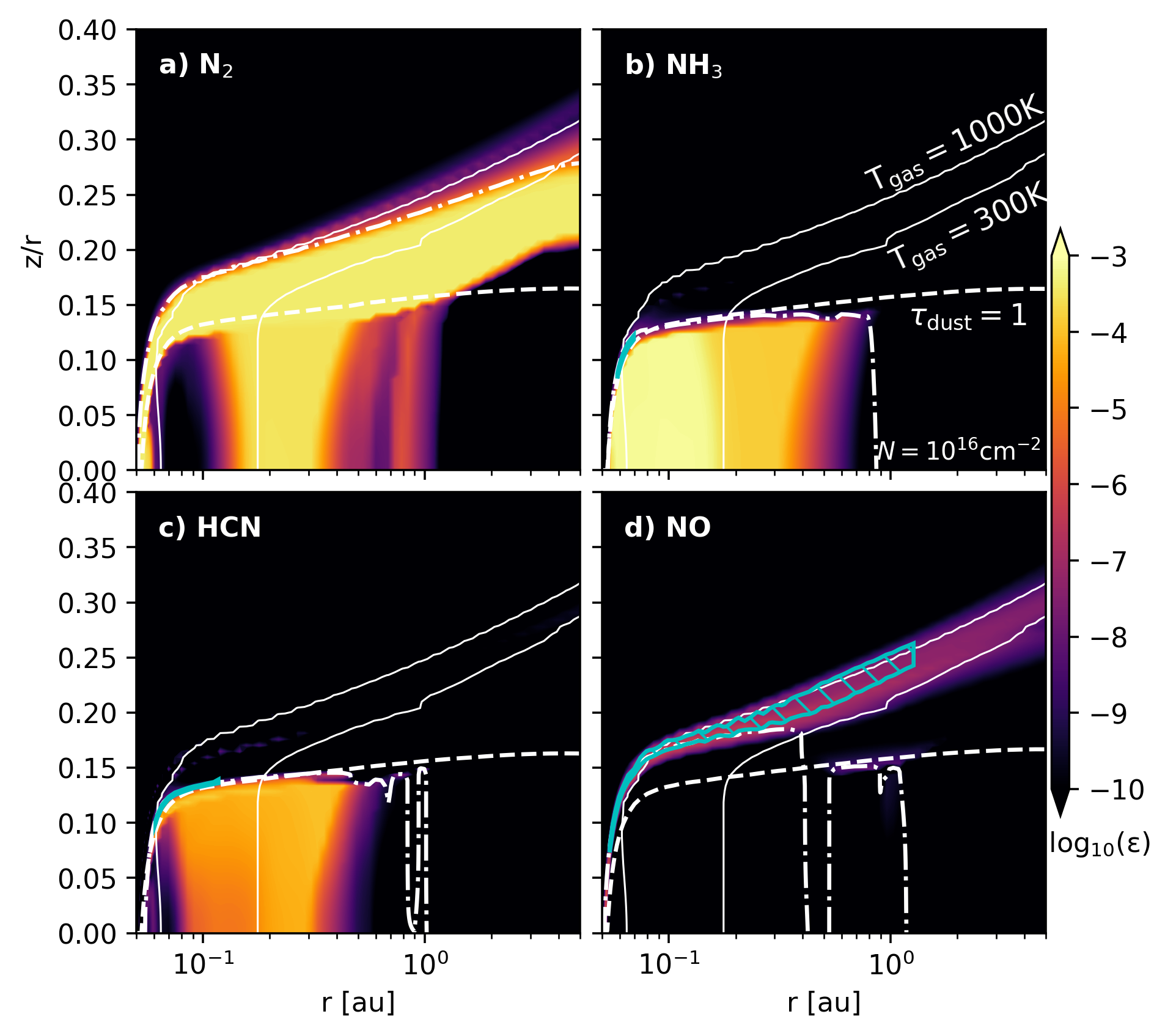}
    \caption{Fractional abundances of major nitrogen-carrier species \ce{N2} (a), \ce{NH3} (b), HCN (c), and NO (d) for the reference C/O=0.45 model with enhanced N/H. The dashed line shows the mean dust optical depth surface ($\tau_{\rm dust}=1$) measured between 4.9-28\,$\mu$m for N$_2$, 5.5-7.0\,$\mu$m for NH$_3$, 13.8-14.5\,$\mu$m for HCN, and 5.1-5.8\,$\mu$m for NO. The emitting regions of HCN, \ce{NH3}, and NO are indicated with blue contours, corresponding to the gas emission at these wavelength ranges. Solid lines show the 1000\,K and 300\,K gas temperature contours. The dash-dot line indicates the vertically integrated column density $N=10^{16}$\,cm$^{-2}$ towards the midplane of the disk.}
    \label{fig:abu_enhN}
\end{figure}

\begin{figure}
    \centering
    \includegraphics[width=0.9\linewidth]{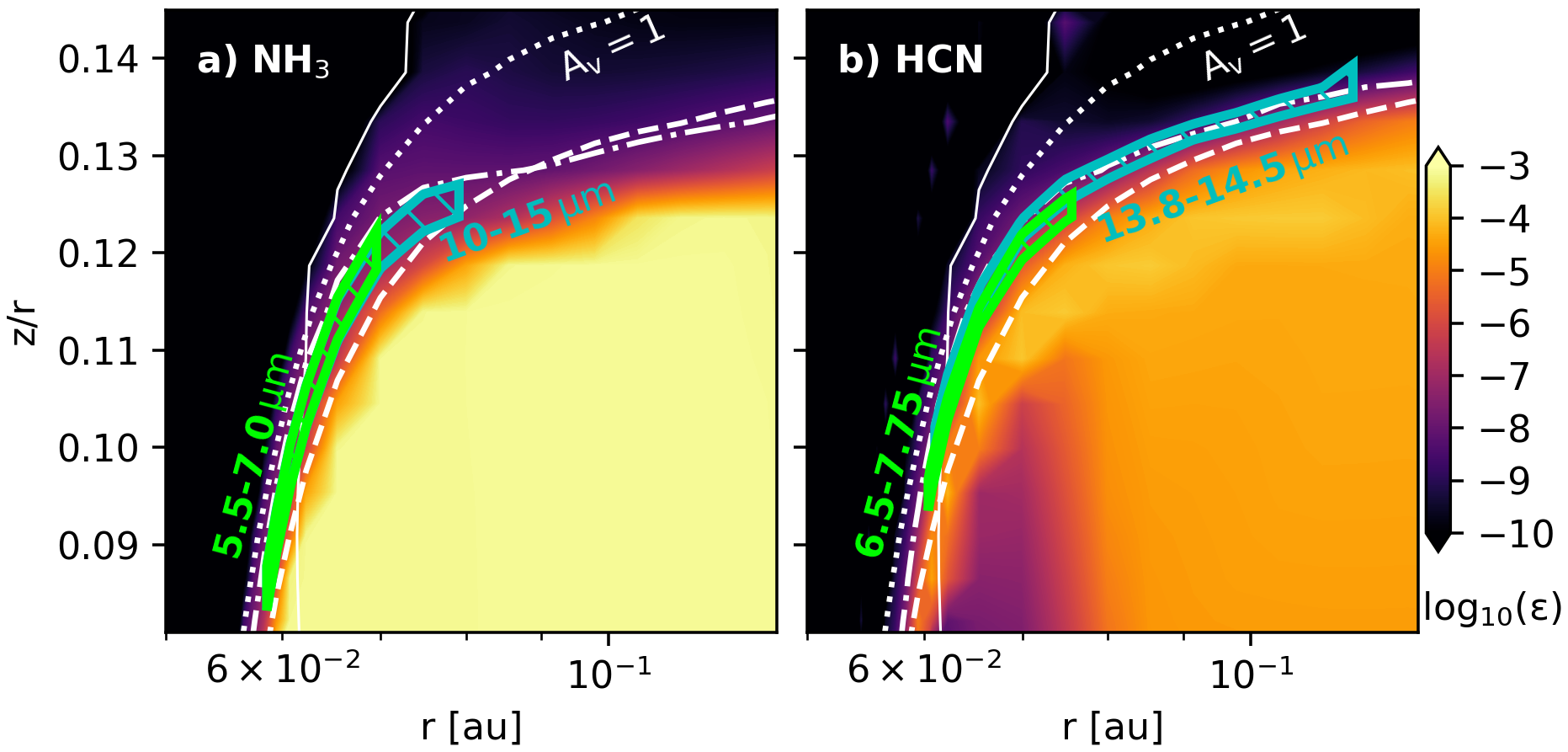}
    \caption{{Zoom-in on the emitting regions of (a) \ce{NH3} and (b) HCN. Colors are same as in Fig.\,\ref{fig:abu_enhN}, with additional lime colored contours indicating the emitting regions of the short wavelength bands at 5.5-7.0\,$\mu$m and 6.5-7.75\,$\mu$m for \ce{NH3} and \ce{HCN} respectively. White dotted lines indicate the A$\rm_v$=1 surface.}}
    \label{fig:abu_enhN_zoom}
\end{figure}

Fig. \ref{fig:abu_enhN} presents the abundance maps of several major N-bearing species (\ce{N2}, \ce{NH3}, HCN, and NO) for the reference, C/O=0.45 model with enhanced N/H. Throughout most of the disk, the gas-phase nitrogen is primarily locked up in \ce{N2}, as its triple bond makes it very stable. In the inner regions, around 0.1 au, \ce{NH3} takes over as the dominant carrier. This reservoir may be difficult to detect, however, as it is largely hidden below the dust continuum. Within the inner 1 au of the disk, HCN is the second most abundant nitrogen carrier. It is located slightly higher up and further out in the disk than \ce{NH3}, and is hence less obscured by dust, which likely accounts for it being readily observed in the IR. {A close-up of the \ce{NH3} and HCN emitting regions, both at short wavelengths ($\sim$6 $\mu$m) and longer wavelengths (10-15 $\mu$m), is presented in Fig. \ref{fig:abu_enhN_zoom}. \citet{heays2017} demonstrate that the photodissociation rate of HCN, assuming a 4000 K blackbody radiation field (which is exactly the same as the input radiation field of our models), is nearly three orders of magnitude smaller than the photodissociation rate of \ce{NH3}. Hence, the \ce{NH3} reservoir is located deeper into the disk than HCN, in regions that are more shielded from the UV radiation.} Finally, we also present the abundance map of NO. This molecule is mostly present in the warm surface layers of the disk. It primarily forms via the \ce{N + OH -> NO + H} reaction \citep{wakelam2012_KIDA}, and therefore its reservoir largely overlaps with that of OH. As this reservoir is located quite high up in the disk, it is the least affected by dust extinction, and thus may be observable despite the relatively low abundance of NO. {The total gas-to-dust ratio is set to 1000 in our models, however due to the settling of the dust, a higher gas-to-dust ratio of $\sim$8000 is reached in the \ce{NH3} and HCN line emitting regions. In the surface layers from which NO emits, the gas-to-dust ratio reaches values as high as $10^5$.} {We note that, for the reference model with solar N/H, the abundance maps of these N-carriers look very similar. The only major difference is that the species have a lower fractional abundance due to the lower nitrogen abundance.}

\begin{figure}
    \centering
    \includegraphics[width=\linewidth]{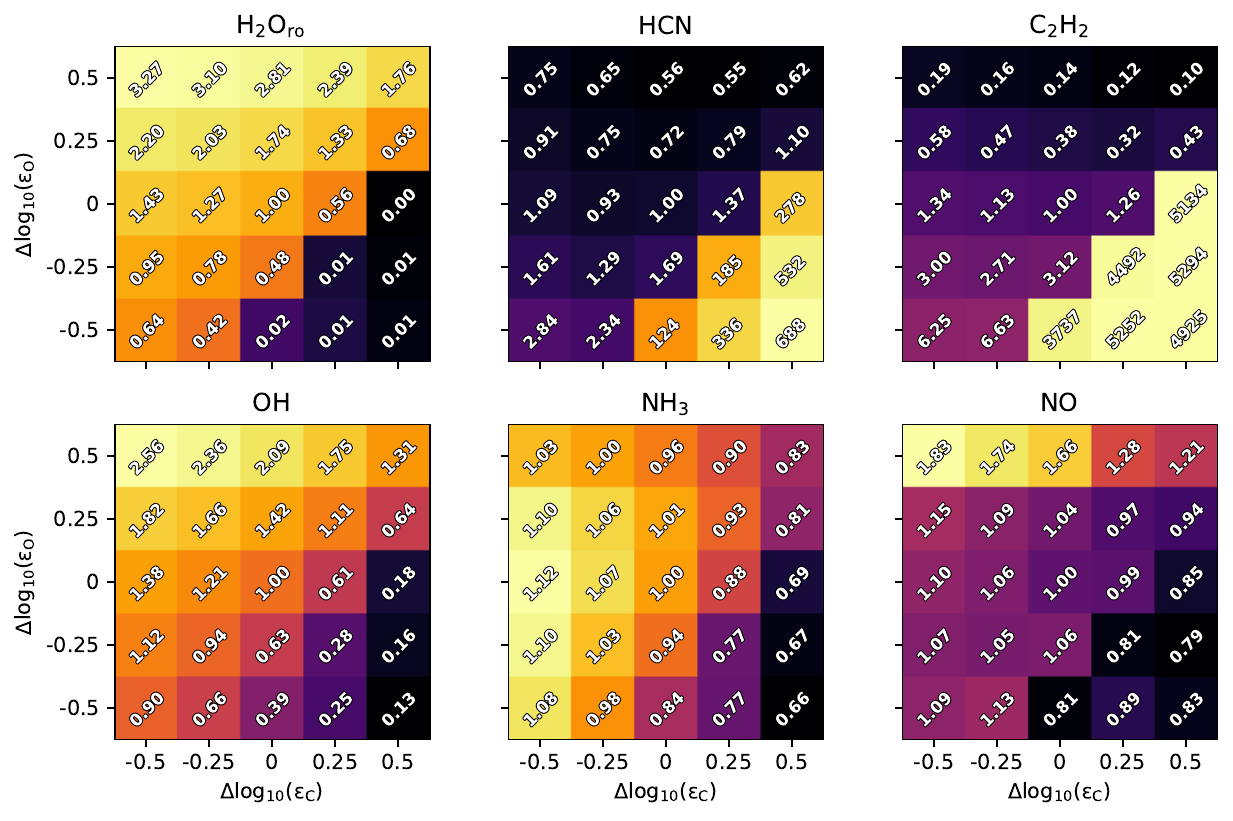}
    \caption{Relative line fluxes of \ce{C2H2}, HCN, \ce{H2O} (rotational), OH, NO, and \ce{NH3} across the grid of models with enhanced N/H, shown as a function of C/H and O/H abundances. The central cell in each panel marks the reference model {with enhanced N/H}; all numbers and colors denote the line flux relative to this model.}
    \label{fig:heatmaps_enhN}
\end{figure}
 
\begin{figure}
    \centering
    \includegraphics[width=0.9\linewidth]{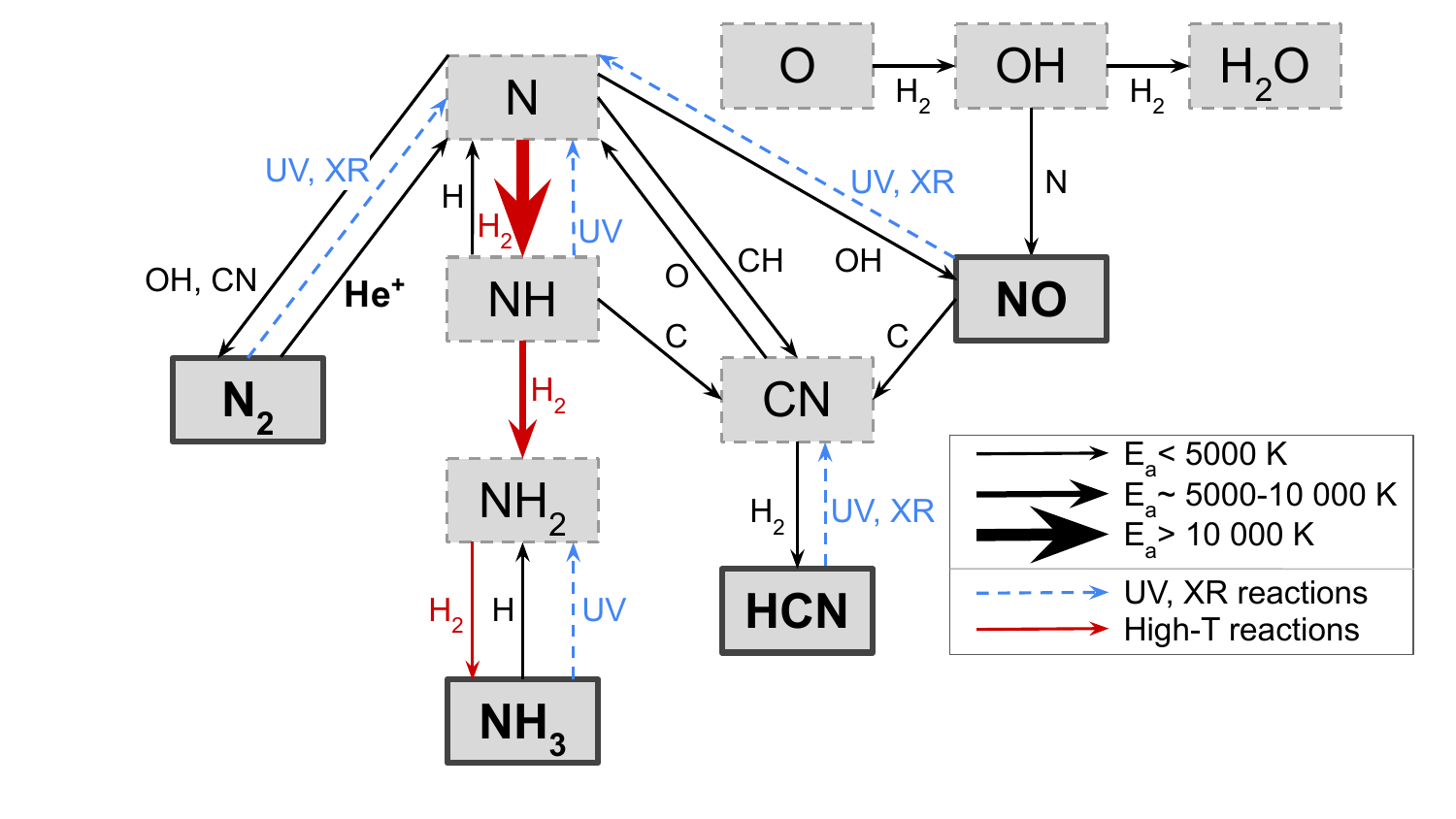}
    \caption{Dominant reactions governing the abundances of the major N-carriers in the inner disk. Adapted from \citet{bast2013, walsh2015} {with a focus on the high-temperature formation routes most relevant for the inner regions of the disk (the low-temperature ion-molecule channels are thus excluded from this graph)}. Gas-phase two-body reactions are indicated with solid lines, and UV or X-ray-induced reactions are indicated with dashed lines. The width of the arrows represents the magnitude of the activation barrier $E_a$. Gas-phase reactions that require high temperatures (several hundred K) are indicated in red.}
    \label{fig:Nchem}
\end{figure}

To demonstrate how the emission of relevant species changes with C/H and O/H, we present the integrated fluxes of \ce{C2H2}, \ce{HCN}, \ce{H2O}, OH, NO, and \ce{NH3} {in the MIRI wavelength range} as a function of C/H and O/H in Fig. \ref{fig:heatmaps_enhN}. {To calculate the integrated fluxes, we adopt the wavelength ranges for \ce{C2H2}, HCN, \ce{H2O}, OH, and \ce{NH3} from \citet{arabhavi2026}, and 5-6 $\mu$m for NO.} The flux of the oxygen-bearing species (\ce{H2O}, OH, and NO) peaks at the highest O/H and lowest C/H (top-left corner of the panel). Gas-phase OH and \ce{H2O} in disks are primarily formed via the \ce{O + H2 -> OH + H} and \ce{OH + H2 -> H2O + H} reactions \citep{charnley1997, vandishoeck2013, walsh2015}. Thus, these species will naturally have a higher abundance in a high-O/H environment, resulting in a higher flux by a factor of 2-3. Additionally, these reactions compete with reactions that lock up gas-phase C and O into CO, which is another major carrier of oxygen in disks. Hence, the fluxes of these species peak at high O/H and low C/H. As previously mentioned, NO is formed primarily from OH, so its flux largely follows the same trend that is seen for OH. An overview of the dominant reactions that govern the abundances of N-bearing molecules in the inner disk is presented in Fig. \ref{fig:Nchem}.

As for the remaining species in Fig. \ref{fig:heatmaps_enhN}, \ce{C2H2} and \ce{HCN} form in abundance when excess C is available \citep{walsh2015, kanwar2024_model, kanwar2024_sz28, colmenares2024}, and thus their fluxes increase sharply by several orders of magnitude when C/O $>$ 1. These two species form primarily via either neutral-neutral or ion-neutral reactions. \ce{C2H2} is formed through reactions starting with either C or \ce{C+} \citep{kanwar2024_model}. HCN forms primarily through the \ce{CN + H2 -> HCN + H} reaction \citep{baulch1994} (Fig. \ref{fig:Nchem}) or through ion-molecule reactions that lead to the formation of \ce{HCN+}. The flux of \ce{NH3} is highest when the C/H is low. This is most likely caused by the competition with HCN at high C/H, as the free nitrogen (that is not already locked up in \ce{N2}) is more likely to become locked up in HCN rather than \ce{NH3}. The \ce{NH3} flux does not show a strong dependence on O/H (flux changes stay at the level of a few percent). There is a slight decrease in \ce{NH3} flux at the highest O/H, likely due to some competition with efficient NO formation in an oxygen-rich environment. {The trends described above also hold for our solar-N/H grid.}

Fig. \ref{fig:Nchem} thus demonstrates that the formation of N-bearing species other than \ce{N2} relies on the release of free atomic nitrogen via the destruction of \ce{N2}, either by photodissociation or by reactions with \ce{He+} \citep{Agundez08,walsh2015}. This atomic nitrogen can then kickstart the nitrogen chemistry of the disk via the \ce{N + H2 -> NH + H} reaction. This reaction is key for the formation of both \ce{NH3} and HCN, and due to its large activation barrier of $12\,650$ K\citep{davidson1990} it acts as a bottleneck for the formation of these species. In the inner regions of the disk, the activation barrier can be overcome by the reaction with vibrationally excited \ce{H2} (H$_2^*$)\citep{tielens1985}. H$_2^*$ can form via FUV pumping in the disk surface, or via collisional excitation in the denser regions closer to the disk midplane.

\begin{figure}
    \centering
    \includegraphics[width=0.6\linewidth]{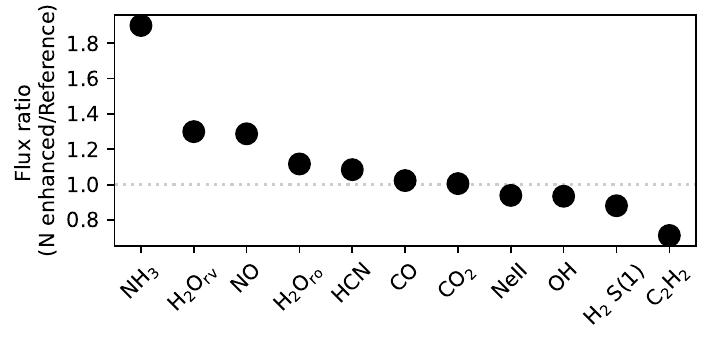}
    \caption{Flux ratios of the reference model with solar N/H and the reference model with enhanced N/H for several relevant species. {The subscripts `ro' and `rv' refer to pure rotational and ro-vibrational line emission, respectively.}}
    \label{fig:ref_vs_enhN}
\end{figure}

It is important to point out that, while the C/O ratio can increase the flux of carbon-bearing species (such as \ce{C2H2} and HCN) by several orders of magnitude (see, e.g., Fig. \ref{fig:heatmaps_enhN}), the N/O ratio does not affect the IR spectrum to the same degree. We compare the flux ratios of several relevant species between our reference model with solar N/H (which has N/O $<$ 1) and our enhanced N/H model (which has N/O $>$ 1) in Fig. \ref{fig:ref_vs_enhN}. All flux ratios remain within a factor 2 of their original values. The \ce{NH3} flux is the most strongly affected, demonstrating an $\sim$90\% increase. The NO flux increases by $\sim$30\%, and the HCN flux remains roughly constant. 

\subsection{Observability of N-carriers}

\begin{figure}
    \centering
    \includegraphics[width=\linewidth]{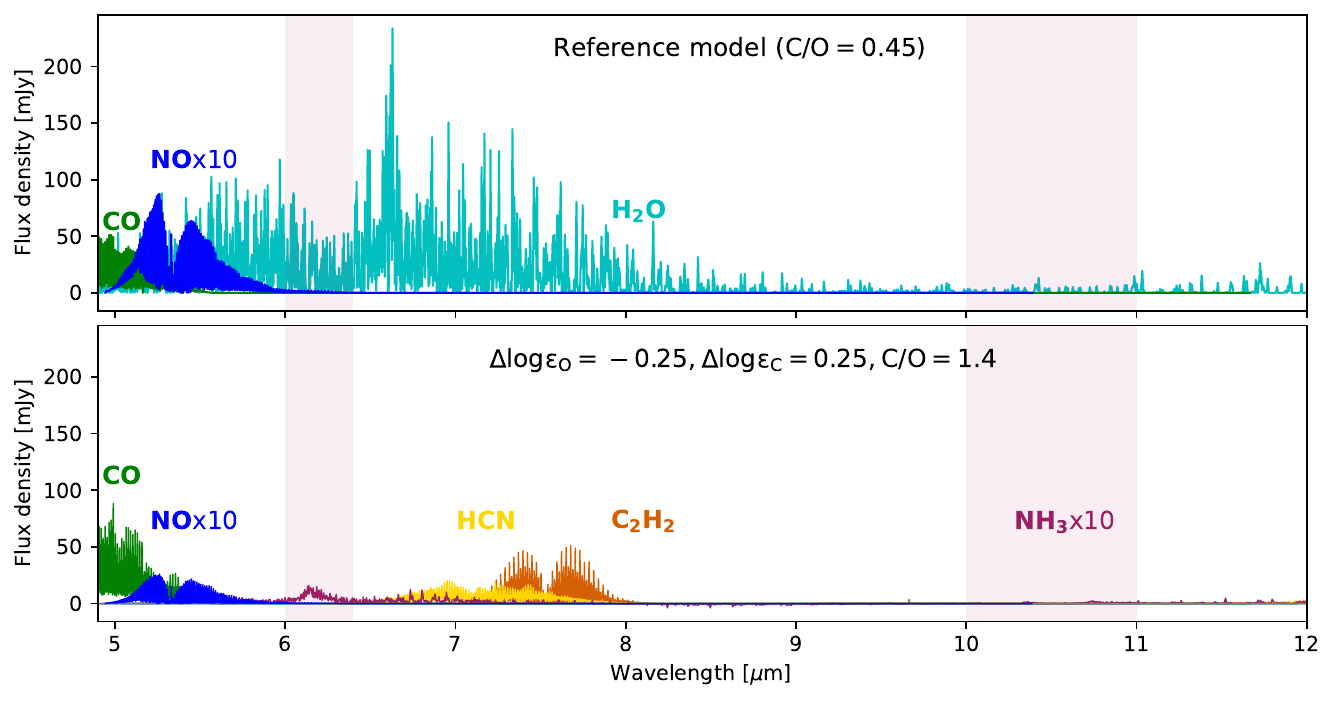}
    \caption{Comparison of the {molecular emission} of a model with C/O $<$ 1  (reference C/H and O/H, enhanced N/H, top panel) and a model with C/O $>$ 1 (C+0.25, O-0.25, enhanced N/H, bottom panel) between 4.9-12.0\,$\mu$m. The contributions from the molecules are indicated in color, respectively. The shaded regions indicate the wavelength ranges with the brightest $\rm NH_3$ emission lines. {NO and \ce{NH3} emission are scaled by a factor 10 for better visualization of the emitting spectral region.}}
    \label{fig:spectra_compare}
\end{figure}

\begin{figure}
    \centering
    \includegraphics[width=\linewidth]{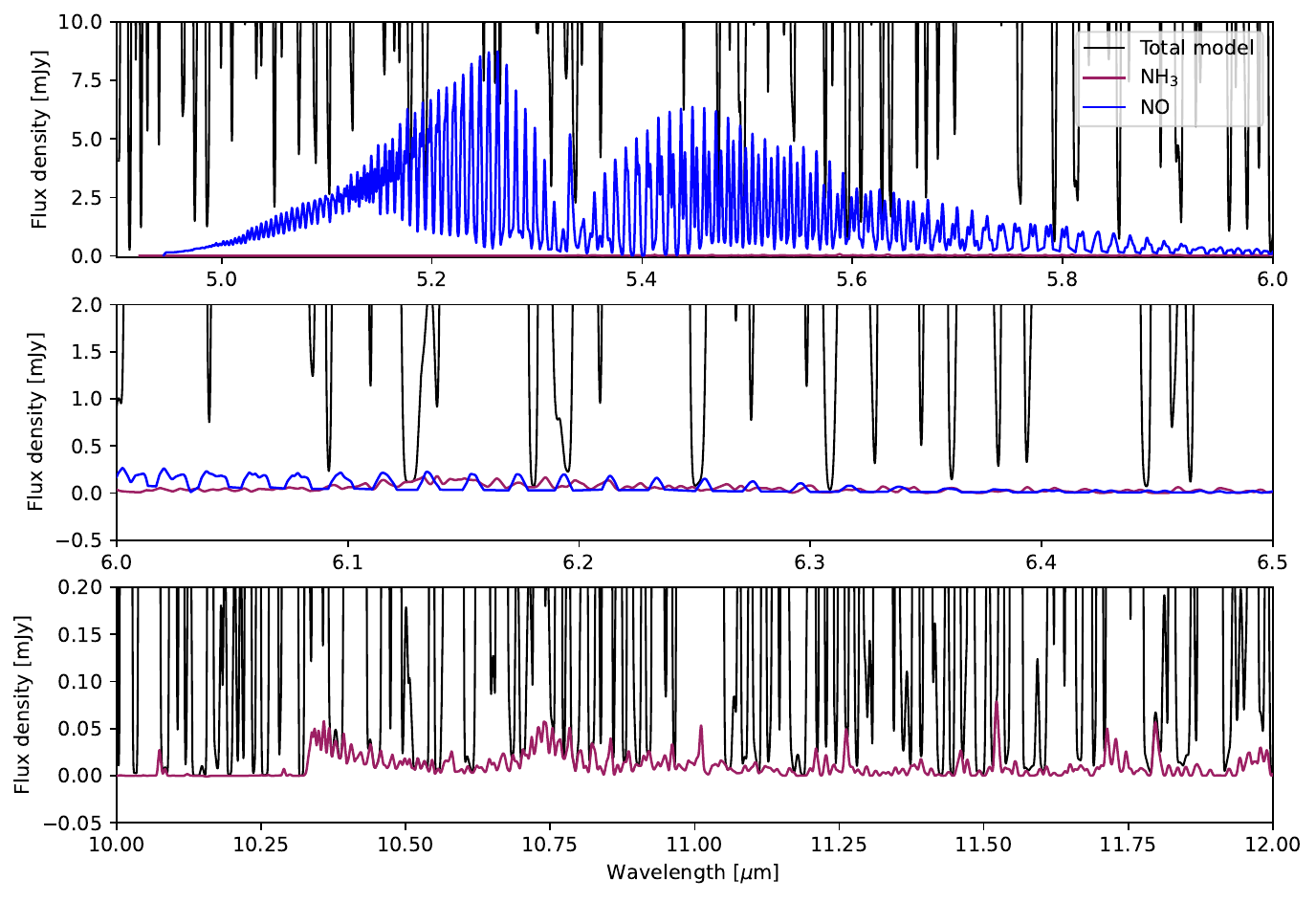}
    \caption{Zoom-ins on the spectra of a C/O $<$ 1 model (reference C/H and O/H, enhanced N/H) between 4.9-6.0, 6.0-6.5, and 10-12 $\mu$m. The full spectrum is shown in black (which is dominated by strong \ce{H2O} and CO emission in this case), and the contributions from NO and \ce{NH3} are indicated in {blue and pink}, respectively. }
    \label{fig:spectra_enhN_CO05}
\end{figure}

\begin{figure}
    \centering
    \includegraphics[width=\linewidth]{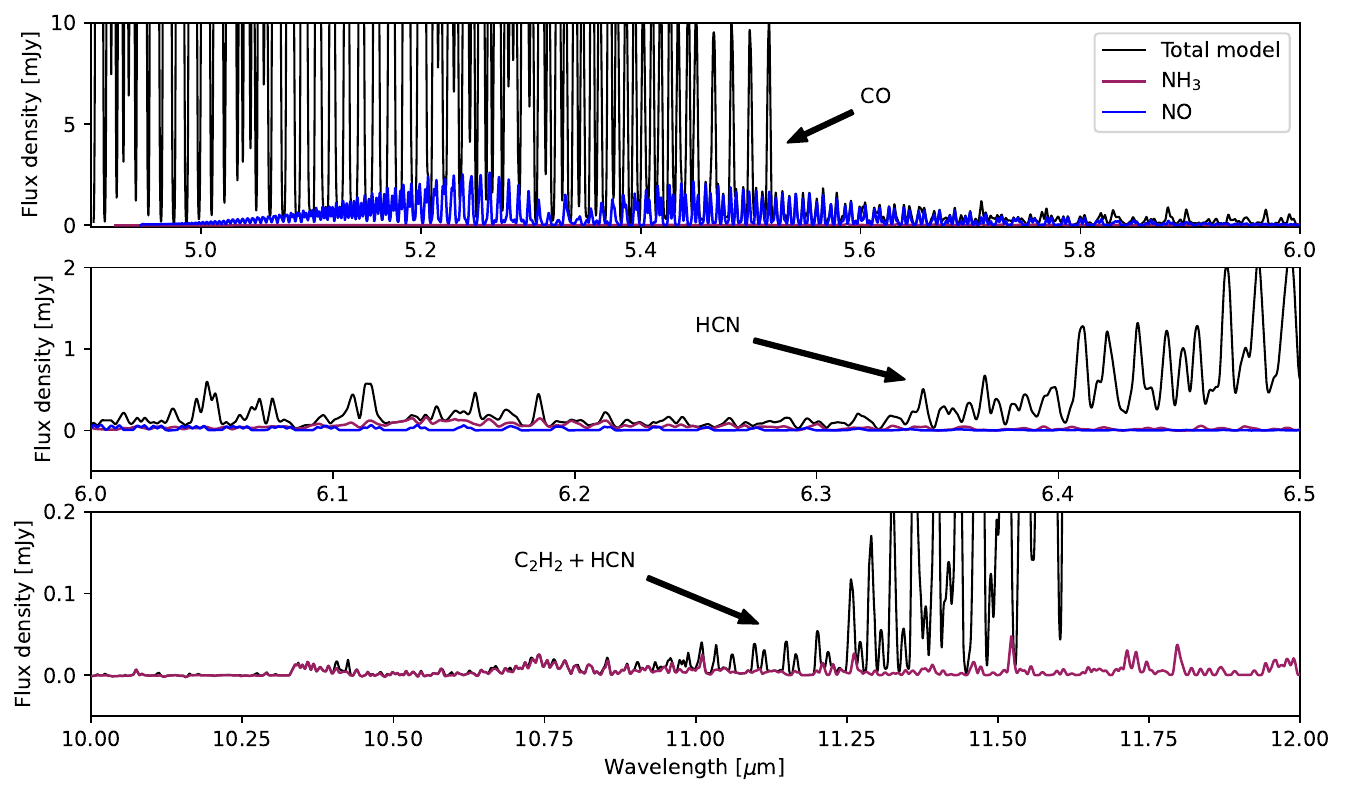}
    \caption{Zoom-ins on the spectra of a C/O $>$ 1 model (C+0.25, O-0.25, enhanced N/H) between 4.9-6.0, 6.0-6.5, and 10-12 $\mu$m. The full spectrum is shown in black, and the contributions from NO and \ce{NH3} are indicated in {blue and pink}, respectively. {We note that the CO spectrum continues beyond 5.5 $\mu$m for higher-$J$ levels than are included in the model.}}
    \label{fig:spectra_enhN_CO1}
\end{figure}

We demonstrate that \ce{NH3} is indeed a major N-carrier in the inner disk, particularly in the extincted regions below the $\tau=1$ surface (Fig. \ref{fig:abu_enhN}), yet its emission has not yet been observed with JWST-MIRI. Hence, we investigate the observability of \ce{NH3} and \ce{NO} with JWST-MIRI. Fig. \ref{fig:spectra_compare} present the full continuum-subtracted IR spectrum between 5-12 $\mu$m at the MIRI spectral resolution for a C/O = 0.45 and a C/O = 1.4 model, both with enhanced N/H. This clearly shows that the spectrum is dominated by O-bearing species when C/O $<$ 1 (mainly \ce{H2O} and CO emission), and it is dominated by C-rich species (\ce{C2H2}, HCN, and CO) when C/O $>$ 1. We present zoom-ins of these spectra in Figs. \ref{fig:spectra_enhN_CO05} and \ref{fig:spectra_enhN_CO1}, which more clearly show the NO and \ce{NH3} emission. We note that these models {are able to reproduce not only} typical observed fluxes of other molecules{, but also, to a reasonable extent, full JWST-MIRI spectra of disks\citep{woitke2024, kanwar2026}}. Therefore, we trust that the models provide reasonable accuracy when it comes to the predicted fluxes for the N-bearing species that have yet to be detected with JWST.

The absolute flux of \ce{NH3}, even with the N/H abundance enhanced by an order of magnitude, is very small in both models (only $\sim$0.3 mJy at most{, within the typical noise level of JWST-MIRI observations}). The emission from this molecule spans from $\sim$6-11 $\mu$m. However, this wavelength range also contains strong \ce{H2O} emission from $\sim$6-8 $\mu$m and beyond 10 $\mu$m (or strong \ce{C2H2} and HCN emission if C/O $>$ 1), all of which make the emission from \ce{NH3} harder to identify due to line blending. Additionally, silicate dust grains also emit strongly between 8-11 $\mu$m. In the models shown in Figs. \ref{fig:spectra_enhN_CO05} and \ref{fig:spectra_enhN_CO1}, this contribution has been subtracted. However, in observations, this contribution can make the detection of \ce{NH3} more difficult. After all, the silicate feature can be quite structured if crystalline silicates are present \citep{jang2025}, and the increased continuum flux can also lead to higher noise levels. Moreover, the dust opacity inside this feature is higher, making the \ce{NH3} gas reservoir more obscured at the wavelengths from which it emits most strongly.

All of this is most evident in the C/O = 0.45 model (Fig. \ref{fig:spectra_enhN_CO05}), where any emission from \ce{NH3} is completely drowned out by \ce{H2O} emission across all wavelength ranges where it is present (see middle and bottom panels of Fig. \ref{fig:spectra_enhN_CO05}). Hence, we find that it is more promising to look for \ce{NH3} in sources with C/O $>$ 1, and {specifically between 6-6.5 $\mu$m or between 10.3-11 $\mu$m.} In the model with C/O = 1.4, {these small windows} are mostly dominated by \ce{NH3} emission, as the \ce{H2O} is now much weaker. Still, the absolute flux between 10-11 $\mu$m only reaches a value of $\sim$0.03 mJy. This is much lower than the typical noise levels achieved for T Tauri disks with JWST-MIRI{, which can range from $\sim$0.3 to a few mJy depending on the strength of the continuum, given a typical signal-to-noise ratio of $\sim$200-450\citep{gasman2025, arulanantham2025}}. At $\sim$6.1 $\mu$m, the emission from \ce{NH3} is slightly stronger at $\sim$0.2 mJy, however this feature is strongly affected by blending with spectral features from HCN, \ce{H2O}, and possibly even NO, as is also the case in our model.

NO emits in the IR between 5-6 $\mu$m (see top panels of Figs. \ref{fig:spectra_enhN_CO05} and \ref{fig:spectra_enhN_CO1}). In our C/O = 0.45 model, its emission is thus completely buried in the high-$J$ P-branch CO lines and \ce{H2O} emission in this wavelength range, even though the NO emission by itself is bright enough to be detected with JWST-MIRI at $\sim$8 mJy. However, in our C/O = 1.4 model, the contribution from \ce{H2O} is almost gone. The CO lines are slightly brighter, however the absolute flux of NO still peaks at an observable 2.5 mJy (and this flux is only $\sim$30\% lower in a solar N/H model, see Fig. \ref{fig:ref_vs_enhN}). As the emission from NO thus still accounts for up to $\sim$25\% of the total emission at wavelengths $\gtrsim$5.5 $\mu$m, careful modeling of the CO emission {(also accounting for the emission of its isotopologues)} could reveal a contribution from NO underneath (we note that the abrupt decrease in CO flux at $\sim$5.5 $\mu$m is a result of an incomplete line list, and is thus not reflective of a real spectrum). 

\subsection{Effects of dust opacity on \ce{NH3} emission}

\begin{figure}
    \centering
    \includegraphics[width=0.9\linewidth]{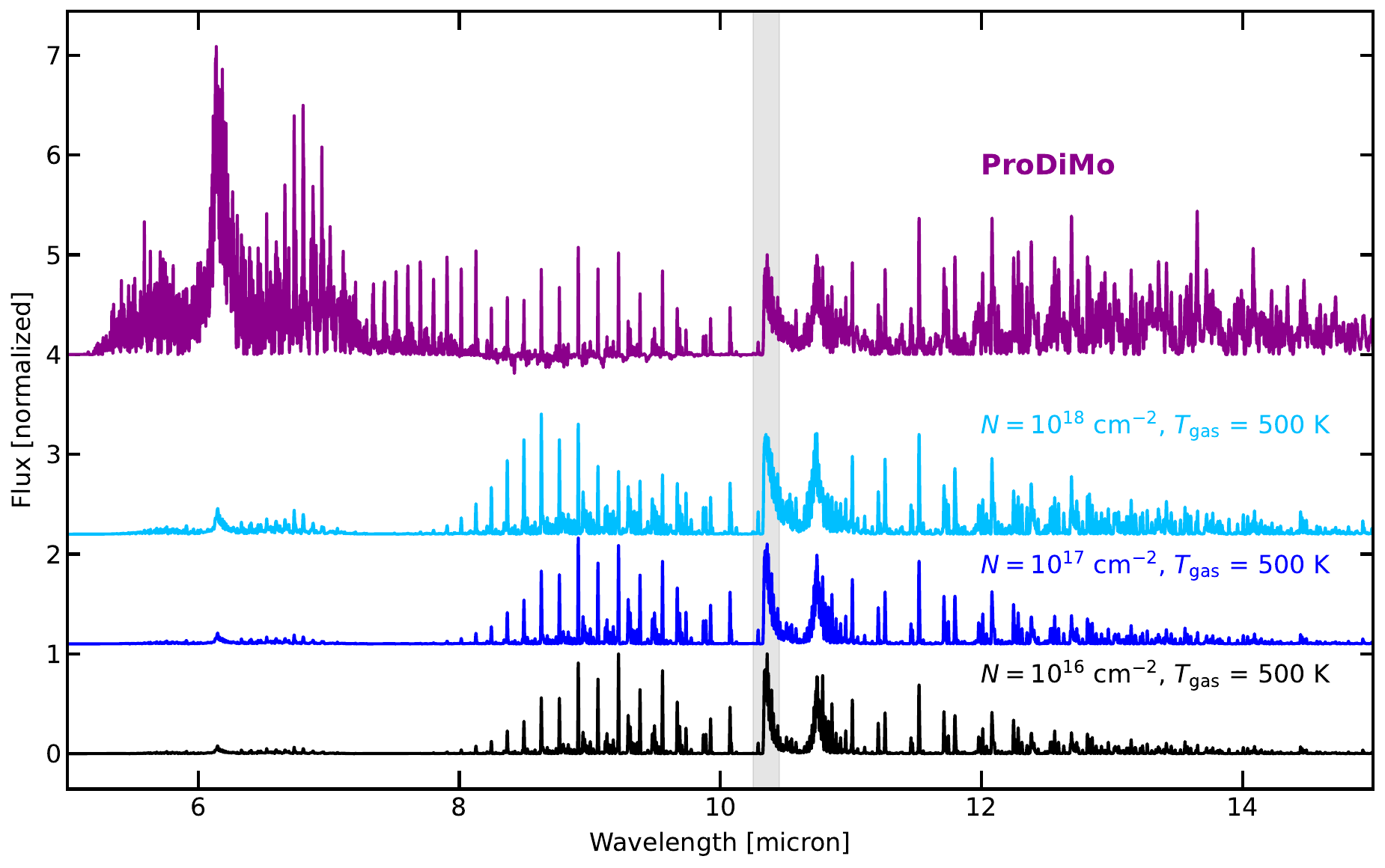}
    \caption{{\ce{NH3} LTE slab models at various temperatures and column densities between 5-15 $\mu$m, compared with the reference ProDiMo model with enhanced N/H. Each spectrum is normalized to the peak flux at 10.5 $\mu$m (indicated in the shaded gray region).}}
    \label{fig:spectra_slabs_ProDiMo}
\end{figure}

To provide further insights into the emission from \ce{NH3}, Fig. \ref{fig:spectra_slabs_ProDiMo} presents several \ce{NH3} LTE slab models, normalized to the flux at 10.5 $\mu$m. {A 500 K model (similar to the gas temperature in the inner disk \ce{NH3} emitting region) is shown at three different column densities, together with the reference ProDiMo \ce{NH3} spectrum with enhanced N/H (same as shown in Fig. \ref{fig:spectra_enhN_CO05}).} 

Interestingly, simple LTE slab models predict the $\nu_2$ \ce{NH3} band at 9-11 $\mu$m to be much stronger than the $\nu_4$ band at 6 $\mu$m, whereas our full disk model finds the opposite. As demonstrated {by} Fig. \ref{fig:spectra_slabs_ProDiMo}, this could be indicative of a {very} high \ce{NH3} column density, as the 6.2 $\mu$m feature grows in relative strength as $N$ increases. Given the high gas-phase abundance in the inner disk ($n$(\ce{NH3})/$n_{\rm H} \gtrsim 10^{-4}$; Fig. \ref{fig:abu_enhN}), the predicted \ce{NH3} column density in our models is indeed high in the inner disk at $\sim10^{18}$ cm$^{-2}$. {However, Fig. \ref{fig:spectra_slabs_ProDiMo} demonstrates that this column density by itself is not enough to create a spectrum that matches our models, as the 6 $\mu$m feature is still weaker than the 9-11 $\mu$m emission. Instead, it is likely that dust plays an important role here. Silicate grains} strongly emit between 9-11 $\mu$m, {which} increases the dust opacity at these wavelengths and therefore {preferentially} obscures the \ce{NH3} emission {at 9-11 $\mu$m}. {This effect is not accounted for in the simple LTE slab models presented in Fig. \ref{fig:spectra_slabs_ProDiMo}, but it is present in our ProDiMo models, and hence we see this discrepancy. }

{This effect may also play a role in the prevailing non-detections of warm \ce{NH3} in T Tauri disks. After all, the lines between 9-11 $\mu$m are predicted by LTE slab models to be the brightest features of \ce{NH3}. They are located in a spectral window that is not very contaminated by emission from \ce{H2O}, theoretically making them a good target to use when searching for \ce{NH3} emission. However, they are most affected by dust obscuration, which seems to lower their fluxes significantly.} Therefore, it effectively hides the most prominent emission features by which \ce{NH3} could be identified in a spectrum. {It is likely that most disks would be affected by this,} given the fact that strong silicate emission is almost ubiquitously seen in disks (however with the notable exception of some carbon-rich low-mass star disks\citep{jang2025}). %Even enhancing the N/H abundance by an order of magnitude does not change this, as we demonstrate that the majority of this excess nitrogen will become locked up in \ce{N2} (Fig. \ref{fig:N_radius}).

{Thus, based on the results from our models, it seems that detecting N-carriers other than HCN will likely prove difficult. Still, it is worth performing a deep search in the data for NO and \ce{NH3}. While the HCN emission is very sensitive to a disk's carbon and oxygen content (Fig. \ref{fig:heatmaps_enhN}), it is the least sensitive to the N/H abundance out of our three N-carriers (Fig. \ref{fig:ref_vs_enhN}). Hence, detections of multiple nitrogen-bearing species may help break degeneracies. Moreover, detecting NO provides insight into the only IR-observable N-bearing tracer of the hot surface layers, where HCN and \ce{NH3} do not have large abundances. Finally, \ce{NH3} emission holds further significance, as it is known to be one of the major ice components\citep{boogert2015}. Thus, comparing the abundance in the gas and ice phases can provide important constraints on ice transport or reprocessing\citep{pontoppidan2019}. }

\subsection{Upper limits on NO and \ce{NH3}}\label{subsec:upper_limits}

We investigate the possible presence of NO and \ce{NH3} emission in three disks observed with JWST as part of the MINDS program: the \ce{H2O}-rich source Sz 98 \citep{gasman2023b}, the \ce{CO2}-rich source GW Lup \citep{grant2023}, and the hydrocarbon-rich disk V1094 Sco (Tabone et al. subm.). These three disks together provide a good indication of the range of chemical diversity that is seen in JWST disk observations, and therefore presumably also span a range of different bulk elemental compositions, just like our models. {By choosing a handful of disks that span a diverse range like this, we can test our model predictions of which circumstances provide the best chances of detecting NO or \ce{NH3}. An overview of their stellar properties, compared to those of our models, is presented in Table \ref{tab:stell_props}.}

\begin{table}[]
    \centering
    \caption{Stellar properties of the three disks and the models presented in this work.}
    \begin{tabular}{l c c c c}
    \hline
    \hline
         Source & $M_{*}$ & $L_{*}$ & $T_{\rm eff}$ & $L_X$ \\
          & [M$_\odot$] & [L$_\odot$] & [K] & [erg/s] \\
         \hline
        GW Lup & 0.41 & 0.33 & 3632 & $2.9 \times 10^{29}$ \\
        Sz 98 & 0.67 & 1.51 & 4060 & $4.4 \times 10^{29}$\\
        V1094 Sco & 0.83 & 1.15 & 4205 & $6.3 \times 10^{29}$ \\
        \hline
        ProDiMo models & 0.7 & 1 & 4000 & $10^{30}$\\
        \hline
    \end{tabular} 
    
    $M_*$, $L_*$, and $T_{\rm eff}$ are taken from \citet{testi2022}. $L_X$ values are taken from \citet{gondoin2006}, \citet{krautter1997}, and \citet{gudel2010}.
    \label{tab:stell_props}
\end{table}

\begin{figure}
    \centering
    \includegraphics[width=\linewidth]{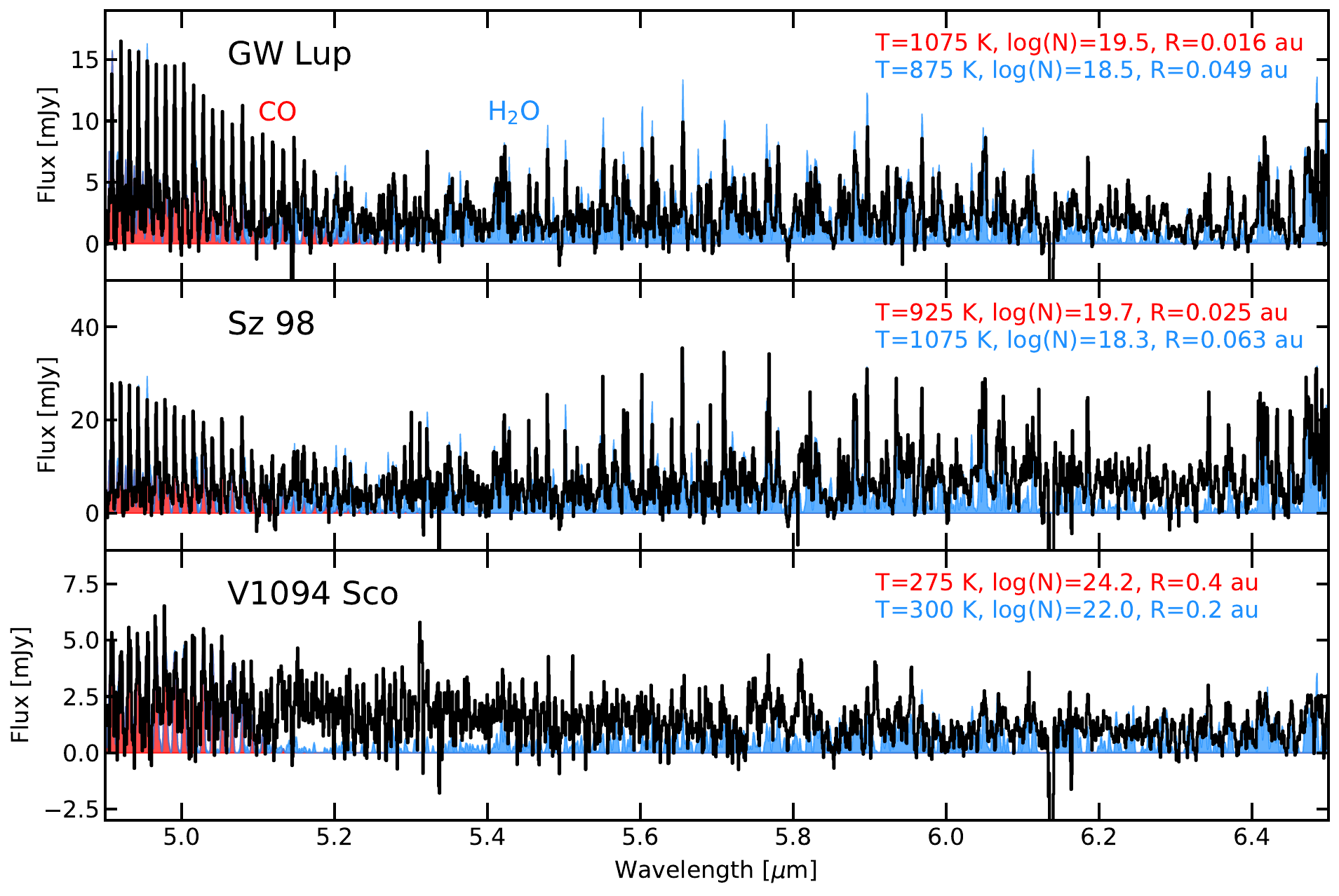}
    \caption{CO and \ce{H2O} LTE slab model fits to GW Lup, Sz 98, and V1094 Sco.}
    \label{fig:slabfits}
\end{figure}

\begin{figure}
    \centering
    \includegraphics[width=\linewidth]{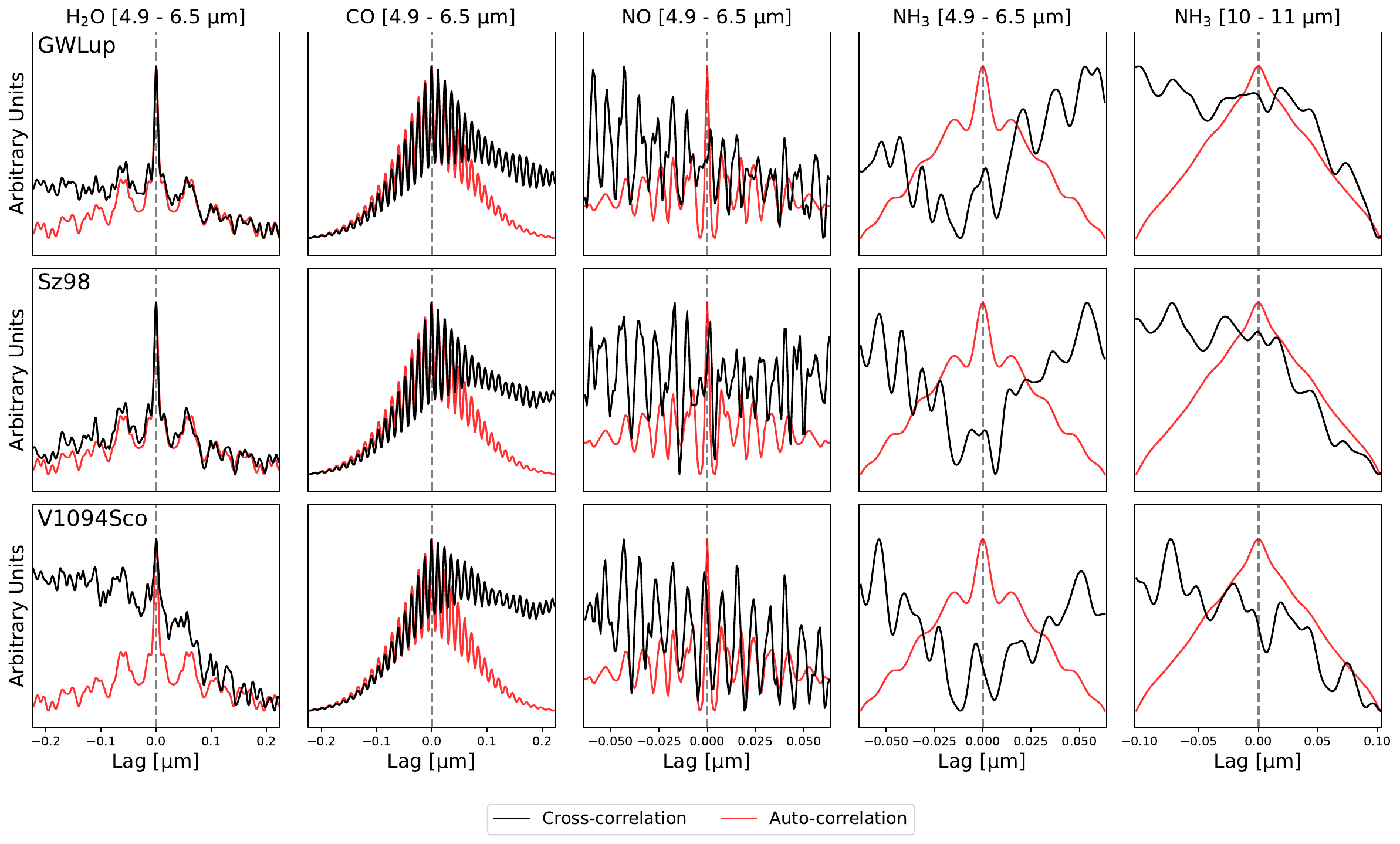}
    \caption{Cross- and auto-correlation signals (black and red lines, respectively) for CO, \ce{H2O}, NO, and \ce{NH3} in GW Lup, Sz 98, and V1094 Sco. The CO, \ce{H2O}, and NO cross-correlations were performed between 4.9 and 6.5 $\mu$m, and the \ce{NH3} cross-correlation was performed between 4.9-6.5 $\mu$m and 10-11 $\mu$m.}
    \label{fig:crosscorr}
\end{figure}

We use a cross-correlation technique to search for detections of NO and \ce{NH3}, as described in \S 2. To this end, we first fit the CO and \ce{H2O} emission between 4.9-6.5 $\mu$m with LTE slab models following \citet{grant2023, tabone2023} (see Fig. \ref{fig:slabfits}). The free parameters in these models are the excitation temperature $T$, column density $N$, and emitting area $\Omega$ which is parameterized as $\Omega=\pi R^2$ where $R$ is the effective emitting radius. In Fig. \ref{fig:crosscorr}, we present the cross-correlation of the data with CO, \ce{H2O}, \ce{NH3}, and NO {template spectra} in black (see Eq. \ref{eq:cc}). For reference, an auto-correlation of the {template spectrum} is also shown in red, representing a pure signal (see Eq. \ref{eq:ac}). For all three sources, clear detections of CO and \ce{H2O} are obtained (Fig. \ref{fig:slabfits}), in line with findings of previous work (Tabone et al. subm., \citep{grant2023, gasman2023b}).
We note that the lack of a single peak in the CO cross- and auto-correlations is due to the periodicity of the emission, combined with the fact that the MIRI wavelength range only covers the high-$J$ tail of the CO $P$-branch, and thus not the full CO emission feature. 

For the NO and \ce{NH3} cross-correlations between 4.9-6.5 $\mu$m, we first subtract the emission from CO and \ce{H2O} from the spectrum using our LTE slab fits. The cross-correlation for \ce{NH3} is performed both between 4.9-6.5 $\mu$m and 10-11 $\mu$m. We do not find a detection in any of our three sources. For NO, we do not find a clear detection in GW Lup and Sz 98, but V1094 Sco reveals a tentative signal (following our criterion described in \S 2). However, we stress that the NO emission has a clear periodicity that lines up relatively well with residual fringing and noise left in the spectrum after the emission from CO and \ce{H2O} has been subtracted. This may lead a cross-correlation to give a false-positive result. Hence, we stress that the possible NO detection in V1094 Sco remains very tentative, and requires further analysis of the data, beyond the scope of this paper. Still, it also highlights the importance of carefully investigating the data for weak, residual emission from less abundant species, as they may provide important constraints.

To derive upper limits on the NO and \ce{NH3} emission, we calculate the $\chi^2$ values for a grid of NO and \ce{NH3} slab models at different values of $T$ and $N$. We scale these slab models to an estimated emitting radius, which we obtain from our reference model ({with enhanced N/H,} see Fig. \ref{fig:abu_enhN}). We estimate the emitting radius of \ce{NH3} to be roughly 0.1 au, and the NO emitting radius to be roughly 1 au. This gives us a likelihood distribution. We then calculate a posterior distribution by multiplying this likelihood distribution with a prior. We set this prior by determining the minimum gas temperature $T_{\rm gas,min}$ within the emitting regions of NO and \ce{NH3} (estimated at 400 and 750 K, respectively), which we also derive from our reference model. We thus use these minimum temperatures to exclude a subset of the slab models. Finally, we draw $10\,000$ samples from our resulting posterior distribution, from which we derive our upper limits as the column density below which 95\% of the values fall. {For \ce{NH3}, we calculate upper limits in two separate wavelength ranges: 4.9-6.5 $\mu$m and 10-11 $\mu$m.}

\begin{table}
  \caption{Derived upper limits for \ce{NH3}.}
  \label{tab:upp_lims_NH3}
  \centering
  \begin{tabular}{l cccccc}
    \hline   
    \hline
    Source & $N$(\ce{NH3}) & $N$(\ce{NH3})  & $T_{\rm gas, min}$ & $R_{\rm em}$ & $N$(\ce{H2O}) & $N$(\ce{NH3})/$N$(\ce{H2O}) \\
     & {4.9-6.5 $\mu$m} & {10-11 $\mu$m} & & & {13.6-16.3 $\mu$m} & \\
      & [cm$^{-2}$] & [cm$^{-2}$] & [K] & [au] & [cm$^{-2}$] &\\ 
    \hline
    GW Lup & $<${5.9} $\times 10^{16}$ & $<${1.4} $\times 10^{15}$ & 750 & 0.1 & $3.2 \times 10^{18}$  & $<0.02$\\
    Sz 98 & $<${1.7} $\times 10^{17}$ & $<${1.2} $\times 10^{16}$  & 750 & 0.1 & $7.9 \times 10^{18}$ & $<${0.02}\\
    V1094 Sco & $<${5.9} $\times 10^{16}$ & $<${2.4} $\times 10^{15}$  & 750 & 0.1 & $1.6 \times 10^{18}$ & $<${0.04}\\
    \hline
  \end{tabular}
\end{table}

% \begin{table}
%   \caption{Derived upper limits for \ce{NH3}.}
%   \label{tab:upp_lims_NH3}
%   \centering
%   \begin{tabular}{l ccccc}
%     \hline   
%     \hline
%     Source & $N$(\ce{NH3})  & $T_{\rm gas, min}$ & $R_{\rm em}$ & $N$(\ce{H2O}) & $N$(\ce{NH3})/$N$(\ce{H2O}) \\
%      & [cm$^{-2}$] & [K] & [au] & [cm$^{-2}$] &\\ 
%     \hline
%     GW Lup & $<${5.9} $\times 10^{16}$ & 750 & 0.1 & $3.2 \times 10^{18}$  & $<0.02$\\
%     Sz 98 & $<${1.7} $\times 10^{17}$  & 750 & 0.1 & $7.9 \times 10^{18}$ & $<${0.02}\\
%     V1094 Sco & $<${5.9} $\times 10^{16}$  & 750 & 0.1 & $1.6 \times 10^{18}$ & $<${0.04}\\
%     \hline
%   \end{tabular}
% \end{table}

\begin{table}
  \caption{Derived upper limits for NO}
  \label{tab:upp_lims_NO}
  \centering
  \begin{tabular}{l ccc}
    \hline
    \hline
    Source & $N$(NO) & $T_{\rm gas, min}$ & $R_{\rm em}$\\
     & {4.9-6.5 $\mu$m} & & \\
    & [cm$^{-2}$] & [K] & [au] \\ 
    \hline
    GW Lup & $<1.0 \times 10^{16}$ & 400 & 1  \\
    Sz 98 & $<2.0\times 10^{16}$ & 400 & 1  \\
    V1094 Sco & $<1.6 \times 10^{16}$ & 400 & 1  \\
    \hline
  \end{tabular}
\end{table}

\begin{figure}
    \centering
    \includegraphics[width=\linewidth]{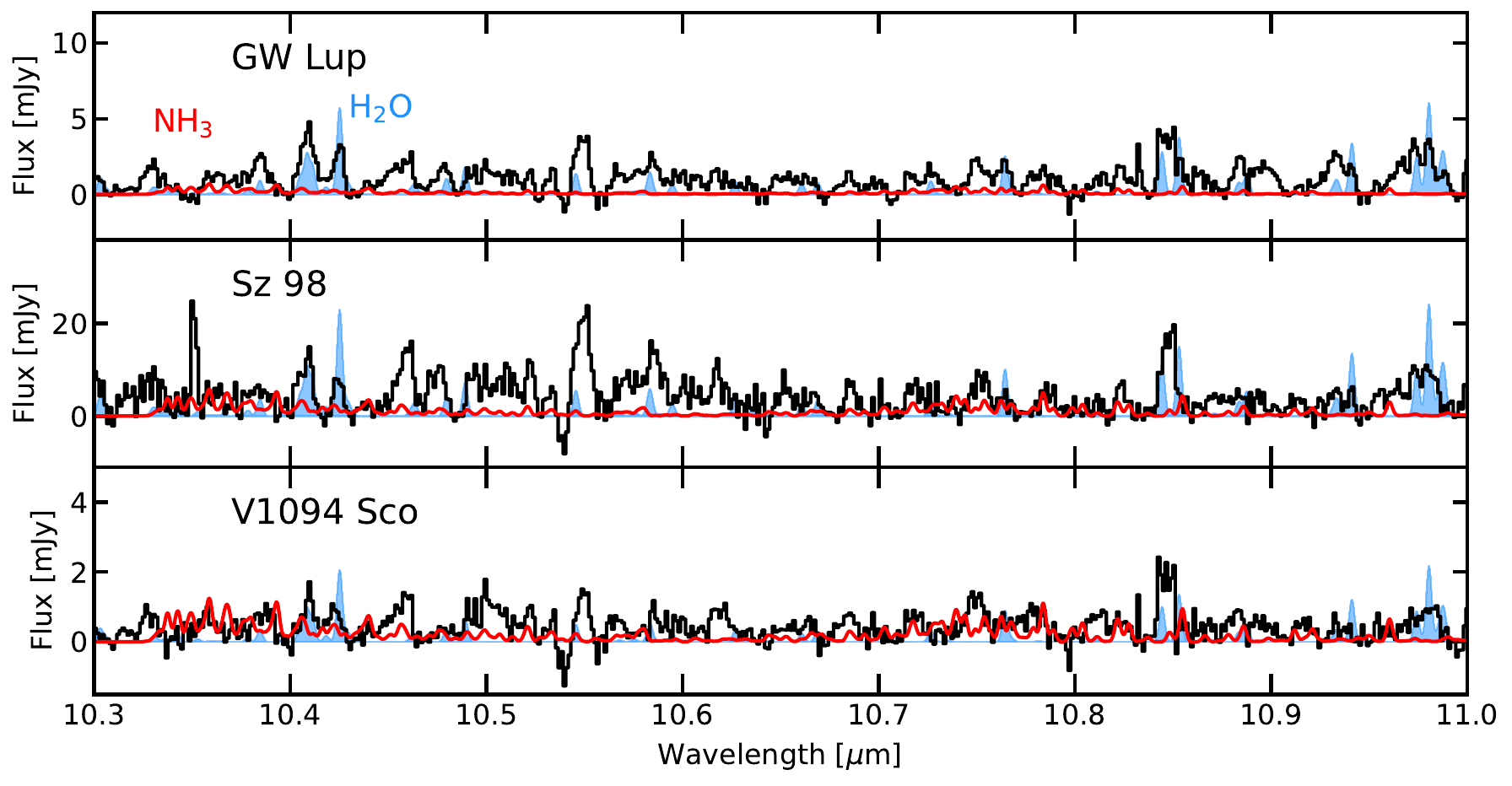}
    \caption{Derived \ce{NH3} upper limits for GW Lup, Sz 98, and V1094 Sco. The parameters of the \ce{NH3} slab model follow Table \ref{tab:upp_lims_NH3}. A scaled $T=600$ K, $N=10^{18}$ cm$^{-2}$ \ce{H2O} slab model is shown in blue.}
    \label{fig:upplims}
\end{figure}

The derived upper limits on \ce{NH3} are given in Table \ref{tab:upp_lims_NH3}{, and Fig. \ref{fig:upplims} presents the upper limits derived at 10-11 $\mu$m together with the data. The derived upper limits at 10-11 $\mu$m are about an order of magnitude lower than those derived at 4.9-6.5 $\mu$m. As Fig. \ref{fig:spectra_slabs_ProDiMo} demonstrates, this makes sense given that the 6 $\mu$m emission from \ce{NH3} is naturally weaker in the LTE slab models. However, these LTE slab models do not account for the effects of dust opacity, as was demonstrated to significantly affect the \ce{NH3} emission at 10-11 $\mu$m as demonstrated by our ProDiMo models. Hence, if this effect were accounted for, a larger column of material could be hidden at these wavelengths, and thus the upper limits derived at 4.9-6.5 $\mu$m may provide a more accurate constraint on the total \ce{NH3} column that is possibly present. }

\begin{figure}
    \centering
    \includegraphics[width=0.7\linewidth]{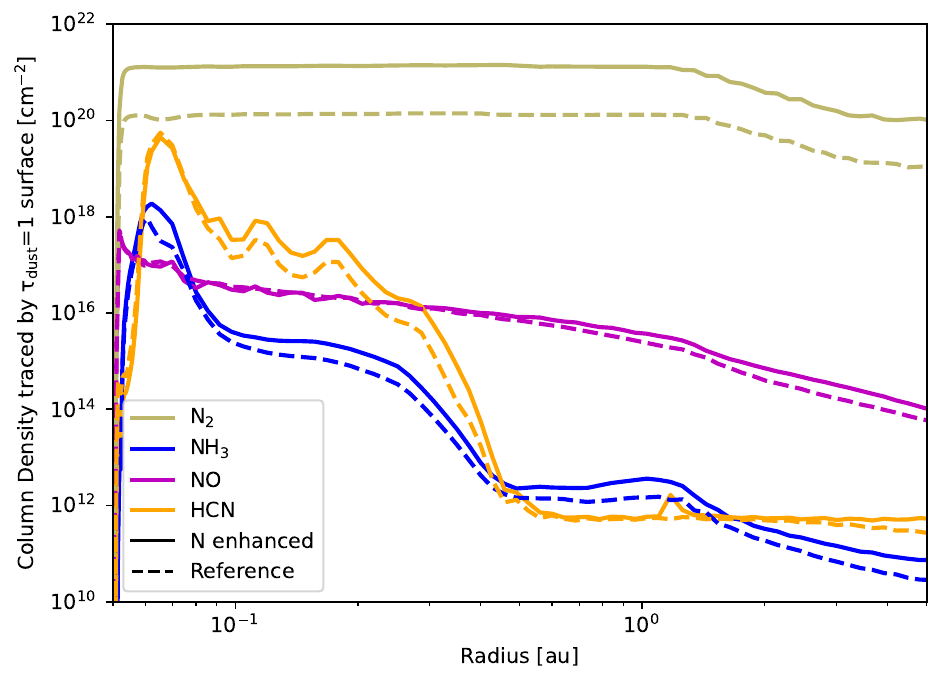}
    \caption{Vertically integrated column density down to the $\tau_{\rm dust}=1$ surface (see Fig. \ref{fig:abu_enhN} caption) as a function of disk radius for several major N-carriers, for both the solar N/H (dashed lines) and enhanced N/H models (solid lines).}
    \label{fig:N_radius}
\end{figure}

{The upper limits derived at 4.9-6.5 $\mu$m} align reasonably well with the predicted \ce{NH3} column density seen above the dust continuum{, which is presented} in Fig. \ref{fig:N_radius}. Though, it demonstrates that we may not be seeing the highest abundance reservoir in the innermost regions of the disk (for which $N$(\ce{NH3}) $\sim10^{18}$ cm$^{-2}$ is predicted). {Alternatively, there may be additional complexity that is not captured by our model. Our upper limits also align quite well with the reported detection of gas-phase \ce{NH3} in absorption in a Class I disk, where a column density of $\sim10^{16}-10^{17}$ cm$^{-2}$ is found\citep{najita2021}. The tentative detection of \ce{NH3} emission in the carbon-rich disk Sz 28 is less consistent with our upper limits, as a much higher column density of $\sim10^{19}$ cm$^{-2}$ is reported\citep{kaeufer2024}. However, \citet{kaeufer2024} also derive a very low temperature ($\lesssim$100 K) for the emission, therefore suggesting that this emission potentially caused by \ce{NH3} may be tracing a different reservoir of gas than the hot \ce{NH3} gas that is expected to be present in the inner disk.}

We also compare {our upper limits derived at 4.9-6.5 $\mu$m} with the column density {of} \ce{H2O}. For this, we take the derived $N$(\ce{H2O}) between 13.6-16.3 $\mu$m in \citet{grant2023, gasman2023b}, and Tabone et al. (subm.). In ices, we know that $N$(\ce{NH3})/$N$(\ce{H2O}) $\sim$6\% \citep{boogert2015}. Given the derived \ce{H2O} column densities, our \ce{NH3} upper limits match quite well with this value. For NO, we present our derived upper limits in Table \ref{tab:upp_lims_NO}. These values align quite well with the NO column density predicted by our models, which is relatively similar across the disk at values between $10^{16}-10^{17}$ cm$^{-2}$.

We note that Fig. \ref{fig:N_radius} also makes it clear why enhancing the elemental N/H abundance by an order of magnitude does not increase the flux of observable N-carriers significantly (see Fig. \ref{fig:ref_vs_enhN}). The column density increase for \ce{NH3}, NO, and HCN between the reference model and enhanced N/H model is minor, whereas the column density of \ce{N2} increases by roughly an order of magnitude. Hence, the vast majority of the enhanced N abundance is locked up in the unobservable \ce{N2}. Unless this \ce{N2} can be more effectively destroyed, for example by increased X-rays or cosmic rays, the nitrogen chemistry of the disk will remain largely unchanged (see also Fig. \ref{fig:Nchem}).

\subsection{Detecting \ce{NH3} in the FIR}

\begin{figure}
    \centering
    \includegraphics[width=0.9\linewidth]{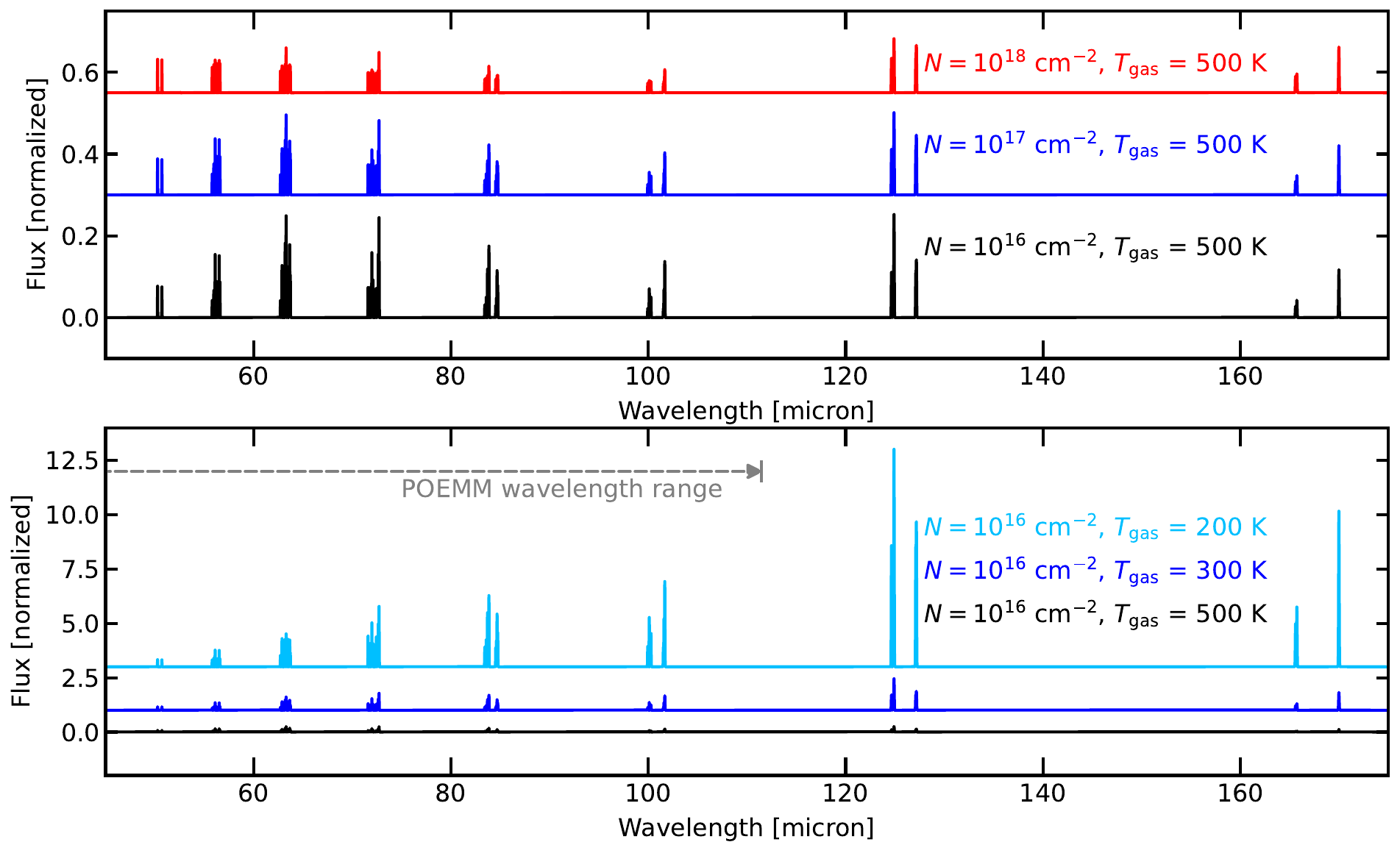}
    \caption{\ce{NH3} LTE slab models at various temperatures and column densities between 45-175 $\mu$m. Each slab model in all panels is normalized to the peak flux at 10.5 $\mu$m ({see Fig. \ref{fig:spectra_slabs_ProDiMo}}).}
    \label{fig:spectra_slabs_FIR}
\end{figure}

Fig. \ref{fig:spectra_slabs_FIR} {demonstrates} possible future prospects for detecting \ce{NH3} emission at longer wavelengths, in the FIR. While the FIR wavelength region is currently unavailable for observations, two future missions are on the horizon that may provide a better characterization of the gas-phase \ce{NH3} content in disks. The Planetary Origins and Evolution Multispectral Monochromator (POEMM) stratospheric balloon mission is a 2 m telescope with a high-resolution ($R = 10^5$) spectrograph that covers wavelengths from 35–112 $\mu$m (see bottom panel of Fig. \ref{fig:spectra_slabs_FIR}). Additionally,  the PRobe far-Infrared Mission for Astrophysics (PRIMA) mission is a 1.8 m space telescope that is currently under development. Its FIRESS
instrument provides spectroscopy over an even broader wavelength range (24–235 $\mu$m) at a superior spectral resolution to JWST/MIRI $(R = 4400 \times 112$ $\mu$m$/\lambda)$, and with a better sensitivity than POEMM. The {top} panel of Fig. \ref{fig:spectra_slabs_FIR} shows the same three slab models as {Fig. \ref{fig:spectra_slabs_ProDiMo}}, still normalized to the flux at 10.5 $\mu$m. This shows that, at a temperature of 500 K, there is a considerable amount of emission in the FIR that could be detected, at a strength of $\sim$20\% of the main feature at 10 $\mu$m. However, it is very likely that the FIR emission does not trace the hot, inner disk gas-phase reservoir of \ce{NH3}, but rather a colder layer further out in the disk (as seen in TW Hya \citep{salinas2016}). Hence, in the bottom panel of Fig. \ref{fig:spectra_slabs_FIR}, we show three LTE slab models with the same column density at three different temperatures (200, 300, and 500 K). This clearly demonstrates that the emission at these long, FIR wavelengths is expected to grow very strongly with decreasing temperature. At the lowest temperature we consider here, the emission at 120 $\mu$m is an order of magnitude stronger than the MIR emission of \ce{NH3}. Hence, while the inner disk gas-phase reservoir of \ce{NH3} will likely remain very hard to observe with current facilities, future facilities in the FIR may provide further insights into the \ce{NH3} budget of disks. 

\section{Discussion}

Our modeling work predicts a high gas-phase \ce{NH3} abundance in the inner disk. However, {our work shows that} this reservoir does not produce emission that is strong enough to be detected with JWST-MIRI. {This is in line with the fact that warm, gas-phase \ce{NH3} remains undetected in T Tauri disks, having only been conclusively detected in a younger Class I source so far\citep{najita2021}. It is worth considering whether the non-detections of \ce{NH3} point to some general process that makes this species so unobservable, which would hence affect all disks equally. For example, the disk chemistry could preferentially lock up the nitrogen into HCN, NO, or \ce{N2} rather than \ce{NH3} in the disk's surface layers that we observe with the IR. We find some evidence for this in our modeling work indeed, as the \ce{NH3} gas-phase reservoir is located slightly deeper down into the disk than HCN (see Fig. \ref{fig:abu_enhN_zoom}). This is likely caused by \ce{NH3} being more easily photodissociated than HCN\citep{heays2017}. Therefore, it can only exist in large amounts in regions of higher dust opacity, below the $\tau=1$ surface, and hence dust would be obscuring its emission.} 

{We note that the difference in photodissociation rates between HCN and \ce{NH3} is quite independent of radiation field temperature (a 4000 K blackbody is considered here), though it weakens slightly for a higher radiation field temperature. \citet{heays2017} demonstrate that the photodissociation rate of HCN is only about 1 order of magnitude lower than that of \ce{NH3} when considering a $10\,000$ K blackbody (which is appropriate for a Herbig star, rather than a T Tauri star), two orders of magnitude less than the difference found for a 4000 K blackbody. Since it is the HCN photodissociation rate that increases in this case, rather than the \ce{NH3} photodissociation rate decreasing, this likely will not make the \ce{NH3} emission more observable. In fact, it may make the HCN emission more difficult to observe, since its reservoir will now also be restricted to regions deeper down in the disk (see also Figure 3 in \citet{walsh2015}). This would be consistent with Herbig disks being typically found to be relatively line-poor, with the exception of emission from OH and \ce{H2O}\citep{pontoppidan2010}.  }

{The obscuration of \ce{NH3} emission by dust further plays a role in its non-detections, as} the brightest emission features of \ce{NH3} are located at 10-11 $\mu$m, which coincides with the wavelength range in which the presence of silicate grains enhance the dust opacity even further. {This may explain} why our models predict the 10-11 $\mu$m feature of \ce{NH3} to be weaker than its 6 $\mu$m feature, which is contrary to what slab models predict (Fig. \ref{fig:spectra_slabs_ProDiMo}). {Moreover, blending with stronger IR emission features, such as those of \ce{H2O}, further complicates any possible detections of \ce{NH3}.} We find that the most unblended features of \ce{NH3} are located around 6 $\mu$m and around 11 $\mu$m in a high-C/O model (Fig. \ref{fig:spectra_enhN_CO1}). Since these features are completely drowned out by \ce{H2O} emission in C/O $<$ 1 models (Fig. \ref{fig:spectra_enhN_CO05}), carbon-rich disks may actually provide {a better chance of detecting the weak features of \ce{NH3}, should they be bright enough to be detected}. This may be corroborated by the tentative detection of \ce{NH3} in the carbon-rich disk Sz 28 \citep{kaeufer2024}{, though we do stress that the emission tentatively detected in this source may trace a different, colder reservoir of \ce{NH3} at large radii rather than the warm, inner reservoir. }The \ce{NH3} emission demonstrates only a weak dependence on the O/H composition of the disk, and a slightly stronger dependence on C/H (Fig. \ref{fig:heatmaps_enhN}). \ce{H2O} emission, on the other hand, is much more sensitive to the O/H. Hence, {if weak features of \ce{NH3} would be present at the noise level, they could be easier to observe} in an inner disk that is strongly oxygen-depleted, but perhaps not too enhanced in carbon, which would strongly increase the emission of HCN. The total \ce{NH3} flux is somewhat sensitive to the N/H elemental ratio (see Fig. \ref{fig:ref_vs_enhN}), yet increasing it by an order of magnitude did not bring the flux to an observable level. {This suggests that observing \ce{NH3} may prove rather difficult, regardless of the elemental composition of the disk. Still, it is worth continuing the search, as expanding this work to a larger sample of disks will still provide very useful upper limits on the \ce{NH3} column density and therefore the \ce{NH3/H2O} column density ratio}.

{We note that we assume a total gas-to-dust ratio of 1000 in this work, rather than the value of 100 that is assumed in \citet{arabhavi2026}, the work on which our grid is based. This choice of increased gas/dust ratio allows us to better match absolute line fluxes observed with JWST-MIRI\citep{woitke2024, meijerink2009}, but we do not find it to have a major impact on our results. As is also known to happen for other species \citep{woitke2018}, the fluxes of NO and \ce{NH3} are increased in our g/d=1000 model compared to a g/d=100 model by a factor of $\sim$5 and $\sim$1.5, respectively. Hence, in a disk with more dust (for example due to a large amount of dust being brought into the inner regions by pebble drift), \ce{NH3} emission may be even more difficult to detect. On the other hand, the emission may be seen somewhat more readily in a disk that is strongly dust-depleted.}

{Additionally, the physical structure of the disk may also affect the predicted line fluxes. The line emitting regions of both \ce{NH3} and HCN are located at the smooth inner rim of the disk (Fig. \ref{fig:abu_enhN_zoom}). \citet{woitke2024} and \citet{henning2024} have demonstrated that the shape of this region can strongly influence IR emission, particularly for species that emit from this region, such as \ce{CH3+}. Additionally, the inner disk radius may also impact molecular fluxes\citep{woitke2018}, and species may be affected differently depending on the location of their gas-phase reservoir in the disk \citep{vlasblom2024}. Hence, this may also influence the observability of N-bearing molecules. }

{Moreover}, we must consider that our modeling work does not include the effects of ice transport. \ce{NH3} ice is known to be present in interstellar ices at an abundance of $\sim$6\% with respect to \ce{H2O}\citep{boogert2015}. If the effects of ice sublimation at the \ce{NH3} ice line (located around $\sim$1 au) were taken into account, the gas-phase \ce{NH3} abundance in the inner regions of the disk may be even further enhanced, and thereby also possibly enhancing its emission. However, the transport of ices occurs mainly in the midplane of the disk. Thus, efficient vertical mixing would be required to transport the released \ce{NH3} gas from the ices into the IR emitting region at $z/r \sim 0.1$, as was also found to be needed for \ce{H2O} \citep{vlasblom2025_H2O}. {However, while such a process may bring the \ce{NH3} gas above the $\tau=1$ surface, where it could emit more strongly, it is likely that such an enhancement would be only temporary. After all, \ce{NH3} will still photodissociate rapidly (on a timescale of only a few years\citep{heays2017}), and therefore a rather large, steady supply of ices may be needed for the \ce{NH3} to remain above the $\tau=1$ surface for any appreciable amount of time. Additionally,} the \ce{NH3} {may also} be chemically converted into other species in the upper layers where chemical timescales are short. Hence, a combined treatment of both ice transport processes and disk chemistry may be needed to obtain further insights into the observability of \ce{NH3} in the planet-forming regions of disks.

Our derived upper limits on \ce{NH3} relative to \ce{H2O} are generally consistent with ice abundances, perhaps hinting at an \ce{NH3} abundance that is a bit lower. This is generally in line with other work. The upper limits on the observable \ce{NH3} abundance in the inner disk derived by \citet{pontoppidan2019} predict an \ce{NH3} abundance almost 10 times lower than the ice abundance. The cold \ce{NH3} detected in the outer disk of TW Hya, on the other hand, was found to have an abundance consistent with an origin directly from ice photodesorption \citep{salinas2016}. This indicates that \ce{NH3} is indeed generally not detected in the expected quantities in the warm surface layers of the inner disk in the mid-infrared, perhaps due to its most abundant reservoir being hidden by the dust. The detection of \ce{NH3} in absorption in the surface layers of the Class I source GV Tau N in \citet{najita2021} can then potentially be attributed to the higher inclination at which the source is viewed. \citet{najita2021} propose that this viewing angle allows them to probe the \ce{NH3} that is located slightly deeper into the disk (as indicated by their derived temperature), that would otherwise be hidden in mid-infrared observations of face-on disks.

We find that the absolute fluxes predicted for the NO emission are observable with JWST-MIRI both when C/O $<$ 1 and C/O $>$ 1. Although, when C/O $<$ 1, strong \ce{H2O} emission will drown out the NO, and hence a carbon-rich disk may provide a better chance of observing NO emission. {A detection of NO in a disk could provide interesting insights, despite the fact that it traces only a relatively small fraction of the total nitrogen budget: namely, it is the only N-carrier observable in the IR that is present in the hot surface layers of the disk, where HCN and \ce{NH3} do not have appreciable gas-phase abundances (see Fig. \ref{fig:abu_enhN}). Hence, despite NO not being found to be very sensitive to the total N/H abundance (Fig. \ref{fig:ref_vs_enhN}), it can give insights into this reservoir that other tracers cannot. }
However, we also note that non-LTE effects may play an important role for the emission of NO, as we know is the case for its parent molecule OH\cite{tabone2021}, which are not accounted for in this work. Aside from JWST-MIRI, the Mid-infrared ELT Imager and Spectrograph on the Extremely Large Telescope (ELT-METIS) may also provide an opportunity to detect the ro-vibrational emission from NO, as the instrument will cover the M band, which contains lines from the high-$J$ R-branch lines from NO around 5 $\mu$m. Since METIS has a much higher spectral resolution than MIRI ($R\approx10^5$ vs. 3000), these lines may be easier to disentangle from the bright CO emission, {especially considering the fact that isotopologues like \ce{^13CO} and \ce{C^18O} also provide a contribution to the total flux}. Hence, ELT-METIS may also provide an interesting opportunity to further characterize the nitrogen budget of disks.

\section{Conclusion}

We present a joint modeling and observational study of \ce{NH3} and NO in the inner regions of protoplanetary disks. Our main conclusions are as follows.
\begin{itemize}
    \item The dominant N-carrier throughout protoplanetary disks is \ce{N2}. In the inner $\sim$1 au, HCN and \ce{NH3} also have abundant gas-phase reservoirs. {\ce{NH3} emits from further in and slightly deeper down into the disk than HCN, as it can only exist in large quantities in regions of the disk that are more shielded by dust. This is a consequence of \ce{NH3} having a larger photodissociation rate than HCN.} NO is present at a lower abundance in the warm surface layers of the disk.
    \item The flux of HCN is strongest when the C/H is high and O/H is low, as it forms through reactions with C-bearing molecules. Conversely, NO forms primarily from OH, and thus its flux is highest when O/H is high and C/H is low. The flux of \ce{NH3} does not show a strong dependence of O/H, but the flux is high when C/H is low due to the competition with HCN.
    \item When a disk has a C/O $>$ 1, this results in a strong increase in abundance and flux of all C-bearing species due to the excess C available. Conversely, a similar effect is not seen with N-bearing species when N/O $>$ 1.
    \item The predicted \ce{NH3} flux from our models is too low to be detectable with JWST-MIRI, which is in line with current reported non-detections. This is likely caused by the \ce{NH3} emission {originating from a region in the disk that is} too extincted by dust to be observed, deep into the disk. {Moreover,} its brightest{, most unblended} emission features {at 10-11 $\mu$m coincide} with the wavelength range where silicate grains enhance the dust opacity {even further, which depresses this emission with respect to shorter wavelength bands}. 
    \item The \ce{NH3} flux remains undetectable even when the N/H elemental abundance is enhanced by an order of magnitude, as the majority of this excess nitrogen is locked up in \ce{N2}.
    \item The predicted NO flux from our models is detectable with MIRI. Its spectral features overlap with the much stronger ro-vibrational CO emission, but these could be disentangled through careful modeling.
    \item We use a cross-correlation technique to look for \ce{NH3} and NO detections in three disks, GW Lup, Sz 98, and V1094 Sco. We do not find a detection of \ce{NH3} in any of the disks, and only V1094 Sco demonstrates a tentative detection of NO. We stress, however, that this could be a false-positive caused by residual fringing and noise in the spectrum, and hence requires further study.
    \item We derive upper limits on the \ce{NH3} and NO column density in three disks, GW Lup, Sz 98, and V1094 Sco. The upper limits on NO align well with the vertically integrated column densities predicted by our models. The upper limits on \ce{NH3} are slightly lower than the peak predicted \ce{NH3} column density, but agree well with predictions of \ce{NH3/H2O} column density ratios in ices.
    \item Future facilities in the FIR may be able to provide further insights into the gas-phase \ce{NH3} budget of disks, as a colder gas-phase reservoir (e.g., released from ices) may be readily observable at wavelengths $>$40 $\mu$m. Additionally, ELT-METIS may provide further opportunities for the detection of NO, as its superior spectral resolution compared with JWST-MIRI would make this emission much easier to disentangle from CO {and its isotopologues}.
\end{itemize}
Our work predicts that, while \ce{NH3} is abundantly present in the inner regions of disks, its emission is likely not detectable with JWST-MIRI. NO, however, provides an interesting alternative for further constraints on the N-budget of disks, as its emission may be detectable. Hence, a deeper search for this molecule in the available data is certainly warranted.

\section*{Acknowledgments}

The authors salute Eric Herbst for his pioneering research elucidating so many aspects of the chemistry of astrochemistry, from cold clouds to warm hot cores and planet-forming disks. E.v.D. is grateful for collaborations and friendship over many decades.  
{We thank the anonymous reviewers for their thoughtful comments that helped improve this manuscript.}
The authors thank the MINDS team for collaborations on JWST data, including those used in this manuscript. 
We also thank Peter Woitke, Christian Rab, and Wing-Fai Thi for their efforts in the development of P{\small RO}D{\small I}M{\small O}.
M.V. and E.v.D. acknowledge support from the ERC grant 101019751 MOLDISK.
A.M.A., I.K, and E.v.D. acknowledge support from grant TOP-1 614.001.751 from the Dutch Research Council (NWO).
E.v.D. also acknowledges support from the Danish National Research Foundation through the Center of Excellence “InterCat” (DNRF150).
B. T. acknowledges the support of the Programme National PCMI of CNRS/INSU with INC/INP cofunded by CEA and CNES.
{The authors declare no competing financial interests.}

%%%%%%%%%%%%%%%%%%%%%%%%%%%%%%%%%%%%%%%%%%%%%%%%%%%%%%%%%%%%%%%%%%%%%
%% If you are using classical BibTeX rather than biblatex,
%% remove the \printbibliography and uncomment the \bibliograpy one
%%%%%%%%%%%%%%%%%%%%%%%%%%%%%%%%%%%%%%%%%%%%%%%%%%%%%%%%%%%%%%%%%%%%%
% \printbibliography
\bibliography{references}

\providecommand{\latin}[1]{#1}
\makeatletter
\providecommand{\doi}
  {\begingroup\let\do\@makeother\dospecials
  \catcode`\{=1 \catcode`\}=2 \doi@aux}
\providecommand{\doi@aux}[1]{\endgroup\texttt{#1}}
\makeatother
\providecommand*\mcitethebibliography{\thebibliography}
\csname @ifundefined\endcsname{endmcitethebibliography}  {\let\endmcitethebibliography\endthebibliography}{}
\begin{mcitethebibliography}{89}
\providecommand*\natexlab[1]{#1}
\providecommand*\mciteSetBstSublistMode[1]{}
\providecommand*\mciteSetBstMaxWidthForm[2]{}
\providecommand*\mciteBstWouldAddEndPuncttrue
  {\def\EndOfBibitem{\unskip.}}
\providecommand*\mciteBstWouldAddEndPunctfalse
  {\let\EndOfBibitem\relax}
\providecommand*\mciteSetBstMidEndSepPunct[3]{}
\providecommand*\mciteSetBstSublistLabelBeginEnd[3]{}
\providecommand*\EndOfBibitem{}
\mciteSetBstSublistMode{f}
\mciteSetBstMaxWidthForm{subitem}{(\alph{mcitesubitemcount})}
\mciteSetBstSublistLabelBeginEnd
  {\mcitemaxwidthsubitemform\space}
  {\relax}
  {\relax}

\bibitem[{{\"O}berg} and {Bergin}(2021){{\"O}berg}, and {Bergin}]{Oberg21}
{{\"O}berg},~K.~I.; {Bergin},~E.~A. {Astrochemistry and compositions of planetary systems}. \emph{\physrep} \textbf{2021}, \emph{893}, 1--48\relax
\mciteBstWouldAddEndPuncttrue
\mciteSetBstMidEndSepPunct{\mcitedefaultmidpunct}
{\mcitedefaultendpunct}{\mcitedefaultseppunct}\relax
\EndOfBibitem
\bibitem[{van Dishoeck}(2006)]{vanDishoeck06}
{van Dishoeck},~E.~F. {Chemistry in low-mass protostellar and protoplanetary regions}. \emph{Proceedings of the National Academy of Science} \textbf{2006}, \emph{103}, 12249--12256\relax
\mciteBstWouldAddEndPuncttrue
\mciteSetBstMidEndSepPunct{\mcitedefaultmidpunct}
{\mcitedefaultendpunct}{\mcitedefaultseppunct}\relax
\EndOfBibitem
\bibitem[{{\"O}berg} \latin{et~al.}(2023){{\"O}berg}, {Facchini}, and {Anderson}]{Oberg23}
{{\"O}berg},~K.~I.; {Facchini},~S.; {Anderson},~D.~E. {Protoplanetary Disk Chemistry}. \emph{\araa} \textbf{2023}, \emph{61}, 287--328\relax
\mciteBstWouldAddEndPuncttrue
\mciteSetBstMidEndSepPunct{\mcitedefaultmidpunct}
{\mcitedefaultendpunct}{\mcitedefaultseppunct}\relax
\EndOfBibitem
\bibitem[{Aikawa} and {Herbst}(1999){Aikawa}, and {Herbst}]{Aikawa99a}
{Aikawa},~Y.; {Herbst},~E. {Molecular evolution in protoplanetary disks. Two-dimensional distributions and column densities of gaseous molecules}. \emph{\aap} \textbf{1999}, \emph{351}, 233--246\relax
\mciteBstWouldAddEndPuncttrue
\mciteSetBstMidEndSepPunct{\mcitedefaultmidpunct}
{\mcitedefaultendpunct}{\mcitedefaultseppunct}\relax
\EndOfBibitem
\bibitem[{Aikawa} \latin{et~al.}(2002){Aikawa}, {van Zadelhoff}, {van Dishoeck}, and {Herbst}]{Aikawa02}
{Aikawa},~Y.; {van Zadelhoff},~G.~J.; {van Dishoeck},~E.~F.; {Herbst},~E. {Warm molecular layers in protoplanetary disks}. \emph{\aap} \textbf{2002}, \emph{386}, 622--632\relax
\mciteBstWouldAddEndPuncttrue
\mciteSetBstMidEndSepPunct{\mcitedefaultmidpunct}
{\mcitedefaultendpunct}{\mcitedefaultseppunct}\relax
\EndOfBibitem
\bibitem[{Herbst}(2014)]{Herbst14}
{Herbst},~E. {Three milieux for interstellar chemistry: gas, dust, and ice}. \emph{Physical Chemistry Chemical Physics (Incorporating Faraday Transactions)} \textbf{2014}, \emph{16}, 3344--3359\relax
\mciteBstWouldAddEndPuncttrue
\mciteSetBstMidEndSepPunct{\mcitedefaultmidpunct}
{\mcitedefaultendpunct}{\mcitedefaultseppunct}\relax
\EndOfBibitem
\bibitem[{McGuire}(2022)]{McGuire22}
{McGuire},~B.~A. {2021 Census of Interstellar, Circumstellar, Extragalactic, Protoplanetary Disk, and Exoplanetary Molecules}. \emph{\apjs} \textbf{2022}, \emph{259}, 30\relax
\mciteBstWouldAddEndPuncttrue
\mciteSetBstMidEndSepPunct{\mcitedefaultmidpunct}
{\mcitedefaultendpunct}{\mcitedefaultseppunct}\relax
\EndOfBibitem
\bibitem[{Booth} \latin{et~al.}(2024){Booth}, {Temmink}, {van Dishoeck}, {Evans}, {Ilee}, {Kama}, {Keyte}, {Law}, {Leemker}, {van der Marel}, {Nomura}, {Notsu}, {{\"O}berg}, and {Walsh}]{Booth24}
{Booth},~A.~S.; {Temmink},~M.; {van Dishoeck},~E.~F.; {Evans},~L.; {Ilee},~J.~D.; {Kama},~M.; {Keyte},~L.; {Law},~C.~J.; {Leemker},~M.; {van der Marel},~N.; {Nomura},~H.; {Notsu},~S.; {{\"O}berg},~K.; {Walsh},~C. {An ALMA Molecular Inventory of Warm Herbig Ae Disks. II. Abundant Complex Organics and Volatile Sulphur in the IRS 48 Disk}. \emph{\aj} \textbf{2024}, \emph{167}, 165\relax
\mciteBstWouldAddEndPuncttrue
\mciteSetBstMidEndSepPunct{\mcitedefaultmidpunct}
{\mcitedefaultendpunct}{\mcitedefaultseppunct}\relax
\EndOfBibitem
\bibitem[{Walsh} \latin{et~al.}(2014){Walsh}, {Millar}, {Nomura}, {Herbst}, {Widicus Weaver}, {Aikawa}, {Laas}, and {Vasyunin}]{Walsh14}
{Walsh},~C.; {Millar},~T.~J.; {Nomura},~H.; {Herbst},~E.; {Widicus Weaver},~S.; {Aikawa},~Y.; {Laas},~J.~C.; {Vasyunin},~A.~I. {Complex organic molecules in protoplanetary disks}. \emph{\aap} \textbf{2014}, \emph{563}, A33\relax
\mciteBstWouldAddEndPuncttrue
\mciteSetBstMidEndSepPunct{\mcitedefaultmidpunct}
{\mcitedefaultendpunct}{\mcitedefaultseppunct}\relax
\EndOfBibitem
\bibitem[{Henning} and {Semenov}(2013){Henning}, and {Semenov}]{henning2013}
{Henning},~T.; {Semenov},~D. {Chemistry in Protoplanetary Disks}. \emph{Chemical Reviews} \textbf{2013}, \emph{113}, 9016--9042\relax
\mciteBstWouldAddEndPuncttrue
\mciteSetBstMidEndSepPunct{\mcitedefaultmidpunct}
{\mcitedefaultendpunct}{\mcitedefaultseppunct}\relax
\EndOfBibitem
\bibitem[{Dawson} and {Johnson}(2018){Dawson}, and {Johnson}]{Dawson18}
{Dawson},~R.~I.; {Johnson},~J.~A. {Origins of Hot Jupiters}. \emph{\araa} \textbf{2018}, \emph{56}, 175--221\relax
\mciteBstWouldAddEndPuncttrue
\mciteSetBstMidEndSepPunct{\mcitedefaultmidpunct}
{\mcitedefaultendpunct}{\mcitedefaultseppunct}\relax
\EndOfBibitem
\bibitem[{Dr{\k{a}}{\.z}kowska} \latin{et~al.}(2023){Dr{\k{a}}{\.z}kowska}, {Bitsch}, {Lambrechts}, {Mulders}, {Harsono}, {Vazan}, {Liu}, {Ormel}, {Kretke}, and {Morbidelli}]{Drazkowska23}
{Dr{\k{a}}{\.z}kowska},~J.; {Bitsch},~B.; {Lambrechts},~M.; {Mulders},~G.~D.; {Harsono},~D.; {Vazan},~A.; {Liu},~B.; {Ormel},~C.~W.; {Kretke},~K.; {Morbidelli},~A. {Planet Formation Theory in the Era of ALMA and Kepler: from Pebbles to Exoplanets}. Protostars and Planets VII. 2023; p 717\relax
\mciteBstWouldAddEndPuncttrue
\mciteSetBstMidEndSepPunct{\mcitedefaultmidpunct}
{\mcitedefaultendpunct}{\mcitedefaultseppunct}\relax
\EndOfBibitem
\bibitem[{Walsh} \latin{et~al.}(2015){Walsh}, {Nomura}, and {van Dishoeck}]{walsh2015}
{Walsh},~C.; {Nomura},~H.; {van Dishoeck},~E. {The molecular composition of the planet-forming regions of protoplanetary disks across the luminosity regime}. \emph{\aap} \textbf{2015}, \emph{582}, A88\relax
\mciteBstWouldAddEndPuncttrue
\mciteSetBstMidEndSepPunct{\mcitedefaultmidpunct}
{\mcitedefaultendpunct}{\mcitedefaultseppunct}\relax
\EndOfBibitem
\bibitem[{Woitke} \latin{et~al.}(2018){Woitke}, {Min}, {Thi}, {Roberts}, {Carmona}, {Kamp}, {M{\'e}nard}, and {Pinte}]{woitke2018}
{Woitke},~P.; {Min},~M.; {Thi},~W.~F.; {Roberts},~C.; {Carmona},~A.; {Kamp},~I.; {M{\'e}nard},~F.; {Pinte},~C. {Modelling mid-infrared molecular emission lines from T Tauri stars}. \emph{\aap} \textbf{2018}, \emph{618}, A57\relax
\mciteBstWouldAddEndPuncttrue
\mciteSetBstMidEndSepPunct{\mcitedefaultmidpunct}
{\mcitedefaultendpunct}{\mcitedefaultseppunct}\relax
\EndOfBibitem
\bibitem[{Rigby} \latin{et~al.}(2023){Rigby}, {Perrin}, {McElwain}, {Kimble}, {Friedman}, {Lallo}, {Doyon}, {Feinberg}, {Ferruit}, {Glasse}, {Rieke}, {Rieke}, {Wright}, {Willott}, {Colon}, {Milam}, {Neff}, {Stark}, {Valenti}, {Abell}, {Abney}, {Abul-Huda}, {Acton}, {Adams}, {Adler}, {Aguilar}, {Ahmed}, {Albert}, {Alberts}, {Aldridge}, {Allen}, {Altenburg}, {{\'A}lvarez-M{\'a}rquez}, {Alves de Oliveira}, {Andersen}, {Anderson}, {Anderson}, {Argyriou}, {Armstrong}, {Arribas}, {Artigau}, {Arvai}, {Atkinson}, and {et al.}]{rigby2023}
{Rigby},~J. \latin{et~al.}  {The Science Performance of JWST as Characterized in Commissioning}. \emph{\pasp} \textbf{2023}, \emph{135}, 048001\relax
\mciteBstWouldAddEndPuncttrue
\mciteSetBstMidEndSepPunct{\mcitedefaultmidpunct}
{\mcitedefaultendpunct}{\mcitedefaultseppunct}\relax
\EndOfBibitem
\bibitem[{Kamp} \latin{et~al.}(2023){Kamp}, {Henning}, {Arabhavi}, {Bettoni}, {Christiaens}, {Gasman}, {Grant}, {Morales-Calder{\'o}n}, {Tabone}, {Abergel}, {Absil}, {Argyriou}, {Barrado}, {Boccaletti}, {Bouwman}, {Caratti o Garatti}, {van Dishoeck}, {Geers}, {Glauser}, {G{\"u}del}, {Guadarrama}, {Jang}, {Kanwar}, {Lagage}, {Lahuis}, {Mueller}, {Nehm{\'e}}, {Olofsson}, {Pantin}, {Pawellek}, {Perotti}, {Ray}, {Rodgers-Lee}, {Samland}, {Scheithauer}, {Schreiber}, {Schwarz}, {Temmink}, {Vandenbussche}, {Vlasblom}, {Waelkens}, {Waters}, and {Wright}]{kamp2023}
{Kamp},~I. \latin{et~al.}  {The chemical inventory of the inner regions of planet-forming disks {\textendash} the JWST/MINDS program}. \emph{Faraday Discussions} \textbf{2023}, \emph{245}, 112--137\relax
\mciteBstWouldAddEndPuncttrue
\mciteSetBstMidEndSepPunct{\mcitedefaultmidpunct}
{\mcitedefaultendpunct}{\mcitedefaultseppunct}\relax
\EndOfBibitem
\bibitem[{van Dishoeck} \latin{et~al.}(2023){van Dishoeck}, {Grant}, {Tabone}, {van Gelder}, {Francis}, {Tychoniec}, {Bettoni}, {Arabhavi}, {Gasman}, {Nazari}, {Vlasblom}, {Kavanagh}, {Christiaens}, {Klaassen}, {Beuther}, {Henning}, and {Kamp}]{vandishoeck2023}
{van Dishoeck},~E.~F. \latin{et~al.}  {The diverse chemistry of protoplanetary disks as revealed by JWST}. \emph{Faraday Discussions} \textbf{2023}, \emph{245}, 52--79\relax
\mciteBstWouldAddEndPuncttrue
\mciteSetBstMidEndSepPunct{\mcitedefaultmidpunct}
{\mcitedefaultendpunct}{\mcitedefaultseppunct}\relax
\EndOfBibitem
\bibitem[{Banzatti} \latin{et~al.}(2023){Banzatti}, {Pontoppidan}, {Carr}, {Jellison}, {Pascucci}, {Najita}, {Mu{\~n}oz-Romero}, {{\"O}berg}, {Kalyaan}, {Pinilla}, {Krijt}, {Long}, {Lambrechts}, {Rosotti}, {Herczeg}, {Salyk}, {Zhang}, {Bergin}, {Ballering}, {Meyer}, {Bruderer}, and {Jdiscs Collaboration}]{banzatti2023}
{Banzatti},~A. \latin{et~al.}  {JWST Reveals Excess Cool Water near the Snow Line in Compact Disks, Consistent with Pebble Drift}. \emph{\apjl} \textbf{2023}, \emph{957}, L22\relax
\mciteBstWouldAddEndPuncttrue
\mciteSetBstMidEndSepPunct{\mcitedefaultmidpunct}
{\mcitedefaultendpunct}{\mcitedefaultseppunct}\relax
\EndOfBibitem
\bibitem[{Pontoppidan} \latin{et~al.}(2024){Pontoppidan}, {Salyk}, {Banzatti}, {Zhang}, {Pascucci}, {{\"O}berg}, {Long}, {Mu{\~n}oz-Romero}, {Carr}, {Najita}, {Blake}, {Arulanantham}, {Andrews}, {Ballering}, {Bergin}, {Calahan}, {Cobb}, {Colmenares}, {Dickson-Vandervelde}, {Dignan}, {Green}, {Heretz}, {Herczeg}, {Kalyaan}, {Krijt}, {Pauly}, {Pinilla}, {Trapman}, and {Xie}]{pontoppidan2024}
{Pontoppidan},~K.~M. \latin{et~al.}  {High-contrast JWST-MIRI Spectroscopy of Planet-forming Disks for the JDISC Survey}. \emph{\apj} \textbf{2024}, \emph{963}, 158\relax
\mciteBstWouldAddEndPuncttrue
\mciteSetBstMidEndSepPunct{\mcitedefaultmidpunct}
{\mcitedefaultendpunct}{\mcitedefaultseppunct}\relax
\EndOfBibitem
\bibitem[{Arulanantham} \latin{et~al.}(2025){Arulanantham}, {Salyk}, {Pontoppidan}, {Banzatti}, {Zhang}, {{\"O}berg}, {Long}, {Carr}, {Najita}, {Pascucci}, {Colmenares}, {Xie}, {Huang}, {Green}, {Andrews}, {Blake}, {Bergin}, {Pinilla}, {Vioque}, {Dahl}, {Raul}, {Krijt}, and {The Jdiscs Collaboration}]{arulanantham2025}
{Arulanantham},~N. \latin{et~al.}  {The JDISC Survey: Linking the Physics and Chemistry of Inner and Outer Protoplanetary Disk Zones}. \emph{\aj} \textbf{2025}, \emph{170}, 67\relax
\mciteBstWouldAddEndPuncttrue
\mciteSetBstMidEndSepPunct{\mcitedefaultmidpunct}
{\mcitedefaultendpunct}{\mcitedefaultseppunct}\relax
\EndOfBibitem
\bibitem[{Pontoppidan} \latin{et~al.}(2014){Pontoppidan}, {Salyk}, {Bergin}, {Brittain}, {Marty}, {Mousis}, and {{\"O}berg}]{Pontoppidan14}
{Pontoppidan},~K.~M.; {Salyk},~C.; {Bergin},~E.~A.; {Brittain},~S.; {Marty},~B.; {Mousis},~O.; {{\"O}berg},~K.~I. {Volatiles in Protoplanetary Disks}. Protostars and Planets VI. 2014; pp 363--385\relax
\mciteBstWouldAddEndPuncttrue
\mciteSetBstMidEndSepPunct{\mcitedefaultmidpunct}
{\mcitedefaultendpunct}{\mcitedefaultseppunct}\relax
\EndOfBibitem
\bibitem[{Wright} \latin{et~al.}(2023){Wright}, {Rieke}, {Glasse}, {Ressler}, {Garc{\'\i}a Mar{\'\i}n}, {Aguilar}, {Alberts}, {{\'A}lvarez-M{\'a}rquez}, {Argyriou}, {Banks}, {Baudoz}, {Boccaletti}, {Bouchet}, {Bouwman}, {Brandl}, {Breda}, {Bright}, {Cale}, {Colina}, {Cossou}, {Coulais}, {Cracraft}, {De Meester}, {Dicken}, {Engesser}, {Etxaluze}, {Fox}, {Friedman}, {Fu}, {Gasman}, {G{\'a}sp{\'a}r}, {Gastaud}, {Geers}, {Glauser}, {Gordon}, {Greene}, {Greve}, {Grundy}, {G{\"u}del}, {Guillard}, {Haderlein}, {Hashimoto}, {Henning}, {Hines}, {Holler}, {Detre}, {Jahromi}, {James}, {Jones}, {Justtanont}, {Kavanagh}, {Kendrew}, {Klaassen}, {Krause}, {Labiano}, {Lagage}, {Lambros}, {Larson}, {Law}, {Lee}, {Libralato}, {Lorenzo Alverez}, {Meixner}, {Morrison}, {Mueller}, {Murray}, {Mycroft}, {Myers}, {Nayak}, {Naylor}, {Nickson}, {Noriega-Crespo}, {{\"O}stlin}, {O'Sullivan}, {Ottens}, {Patapis}, {Penanen}, {Pietraszkiewicz}, {Ray}, {Regan}, {Roteliuk}, {Royer}, {Samara-Ratna}, {Samuelson}, {Sargent}, {Scheithauer},
  {Schneider}, {Schreiber}, {Shaughnessy}, {Sheehan}, {Shivaei}, {Sloan}, {Tamas}, {Teague}, {Temim}, {Tikkanen}, {Tustain}, {van Dishoeck}, {Vandenbussche}, {Weilert}, {Whitehouse}, and {Wolff}]{Wright23}
{Wright},~G.~S. \latin{et~al.}  {The Mid-infrared Instrument for JWST and Its In-flight Performance}. \emph{\pasp} \textbf{2023}, \emph{135}, 048003\relax
\mciteBstWouldAddEndPuncttrue
\mciteSetBstMidEndSepPunct{\mcitedefaultmidpunct}
{\mcitedefaultendpunct}{\mcitedefaultseppunct}\relax
\EndOfBibitem
\bibitem[{Temmink} \latin{et~al.}(2025){Temmink}, {Sellek}, {Gasman}, {van Dishoeck}, {Vlasblom}, {Pranger}, {G{\"u}del}, {Henning}, {Lagage}, {Caratti o Garatti}, {Kamp}, {Olofsson}, {Arabhavi}, {Grant}, {Kaeufer}, {Kurtovic}, {Perotti}, {Samland}, {Schwarz}, and {Tabone}]{temmink2025}
{Temmink},~M. \latin{et~al.}  {MINDS: Water reservoirs of compact planet-forming dust discs: A diversity of H$_{2}$O distributions}. \emph{\aap} \textbf{2025}, \emph{699}, A134\relax
\mciteBstWouldAddEndPuncttrue
\mciteSetBstMidEndSepPunct{\mcitedefaultmidpunct}
{\mcitedefaultendpunct}{\mcitedefaultseppunct}\relax
\EndOfBibitem
\bibitem[{Banzatti} \latin{et~al.}(2025){Banzatti}, {Salyk}, {Pontoppidan}, {Carr}, {Zhang}, {Arulanantham}, {Krijt}, {{\"O}berg}, {Cleeves}, {Najita}, {Pascucci}, {Blake}, {Romero-Mirza}, {Bergin}, {Cieza}, {Pinilla}, {Long}, {Mallaney}, {Xie}, {Waggoner}, {Kaeufer}, and {The Jdiscs Collaboration}]{banzatti2025}
{Banzatti},~A. \latin{et~al.}  {Water in Protoplanetary Disks with JWST-MIRI: Spectral Excitation Atlas and Radial Distribution from Temperature Diagnostic Diagrams and Doppler Mapping}. \emph{\aj} \textbf{2025}, \emph{169}, 165\relax
\mciteBstWouldAddEndPuncttrue
\mciteSetBstMidEndSepPunct{\mcitedefaultmidpunct}
{\mcitedefaultendpunct}{\mcitedefaultseppunct}\relax
\EndOfBibitem
\bibitem[{Tabone} \latin{et~al.}(2023){Tabone}, {Bettoni}, {van Dishoeck}, {Arabhavi}, {Grant}, {Gasman}, {Henning}, {Kamp}, {G{\"u}del}, {Lagage}, {Ray}, {Vandenbussche}, {Abergel}, {Absil}, {Argyriou}, {Barrado}, {Boccaletti}, {Bouwman}, {Garatti}, {Geers}, {Glauser}, {Justannont}, {Lahuis}, {Mueller}, {Nehm{\'e}}, {Olofsson}, {Pantin}, {Scheithauer}, {Waelkens}, {Waters}, {Black}, {Christiaens}, {Guadarrama}, {Morales-Calder{\'o}n}, {Jang}, {Kanwar}, {Pawellek}, {Perotti}, {Perrin}, {Rodgers-Lee}, {Samland}, {Schreiber}, {Schwarz}, {Colina}, {{\"O}stlin}, and {Wright}]{tabone2023}
{Tabone},~B. \latin{et~al.}  {A rich hydrocarbon chemistry and high C to O ratio in the inner disk around a very low-mass star}. \emph{Nature Astronomy} \textbf{2023}, \relax
\mciteBstWouldAddEndPunctfalse
\mciteSetBstMidEndSepPunct{\mcitedefaultmidpunct}
{}{\mcitedefaultseppunct}\relax
\EndOfBibitem
\bibitem[{Arabhavi} \latin{et~al.}(2024){Arabhavi}, {Kamp}, {Henning}, {van Dishoeck}, {Christiaens}, {Gasman}, {Perrin}, {G{\"u}del}, {Tabone}, {Kanwar}, {Waters}, {Pascucci}, {Samland}, {Perotti}, {Bettoni}, {Grant}, {Lagage}, {Ray}, {Vandenbussche}, {Absil}, {Argyriou}, {Barrado}, {Boccaletti}, {Bouwman}, {Caratti o Garatti}, {Glauser}, {Lahuis}, {Mueller}, {Olofsson}, {Pantin}, {Scheithauer}, {Morales-Calder{\'o}n}, {Franceschi}, {Jang}, {Pawellek}, {Rodgers-Lee}, {Schreiber}, {Schwarz}, {Temmink}, {Vlasblom}, {Wright}, {Colina}, and {{\"O}stlin}]{arabhavi2024}
{Arabhavi},~A.~M. \latin{et~al.}  {Abundant hydrocarbons in the disk around a very-low-mass star}. \emph{Science} \textbf{2024}, \emph{384}, 1086--1090\relax
\mciteBstWouldAddEndPuncttrue
\mciteSetBstMidEndSepPunct{\mcitedefaultmidpunct}
{\mcitedefaultendpunct}{\mcitedefaultseppunct}\relax
\EndOfBibitem
\bibitem[{Arabhavi} \latin{et~al.}(2025){Arabhavi}, {Kamp}, {Henning}, {van Dishoeck}, {Jang}, {Waters}, {Christiaens}, {Gasman}, {Pascucci}, {Perotti}, {Grant}, {G{\"u}del}, {Lagage}, {Barrado}, {Caratti o Garatti}, {Lahuis}, {Kaeufer}, {Kanwar}, {Morales-Calder{\'o}n}, {Schwarz}, {Sellek}, {Tabone}, {Temmink}, {Vlasblom}, and {Patapis}]{Arabhavi25a}
{Arabhavi},~A.~M. \latin{et~al.}  {MINDS: The very low-mass star and brown dwarf sample: Detections and trends in the inner disk gas}. \emph{\aap} \textbf{2025}, \emph{699}, A194\relax
\mciteBstWouldAddEndPuncttrue
\mciteSetBstMidEndSepPunct{\mcitedefaultmidpunct}
{\mcitedefaultendpunct}{\mcitedefaultseppunct}\relax
\EndOfBibitem
\bibitem[{Long} \latin{et~al.}(2025){Long}, {Pascucci}, {Houge}, {Banzatti}, {Pontoppidan}, {Najita}, {Krijt}, {Xie}, {Williams}, {Herczeg}, {Andrews}, {Bergin}, {Blake}, {Colmenares}, {Harsono}, {Romero-Mirza}, {Li}, {Lu}, {Pinilla}, {Wilner}, {Vioque}, {Zhang}, and {JDISCS Collaboration}]{Long25}
{Long},~F. \latin{et~al.}  {The First JWST View of a 30-Myr-old Protoplanetary Disk Reveals a Late-stage Carbon-rich Phase}. \emph{\apjl} \textbf{2025}, \emph{978}, L30\relax
\mciteBstWouldAddEndPuncttrue
\mciteSetBstMidEndSepPunct{\mcitedefaultmidpunct}
{\mcitedefaultendpunct}{\mcitedefaultseppunct}\relax
\EndOfBibitem
\bibitem[{Kanwar} \latin{et~al.}(2024){Kanwar}, {Kamp}, {Jang}, {Waters}, {van Dishoeck}, {Christiaens}, {Arabhavi}, {Henning}, {G{\"u}del}, {Woitke}, {Absil}, {Barrado}, {Caratti o Garatti}, {Glauser}, {Lahuis}, {Scheithauer}, {Vandenbussche}, {Gasman}, {Grant}, {Kurtovic}, {Perotti}, {Tabone}, and {Temmink}]{kanwar2024_sz28}
{Kanwar},~J. \latin{et~al.}  {MINDS. Hydrocarbons detected by JWST/MIRI in the inner disk of Sz28 consistent with a high C/O gas-phase chemistry}. \emph{\aap} \textbf{2024}, \emph{689}, A231\relax
\mciteBstWouldAddEndPuncttrue
\mciteSetBstMidEndSepPunct{\mcitedefaultmidpunct}
{\mcitedefaultendpunct}{\mcitedefaultseppunct}\relax
\EndOfBibitem
\bibitem[{Colmenares} \latin{et~al.}(2024){Colmenares}, {Bergin}, {Salyk}, {Pontoppidan}, {Arulanantham}, {Calahan}, {Banzatti}, {Andrews}, {Blake}, {Ciesla}, {Green}, {Long}, {Lambrechts}, {Najita}, {Pascucci}, {Pinilla}, {Krijt}, {Trapman}, and {Jdiscs Collaboration}]{colmenares2024}
{Colmenares},~M.~J. \latin{et~al.}  {JWST/MIRI Detection of a Carbon-rich Chemistry in the Disk of a Solar Nebula Analog}. \emph{\apj} \textbf{2024}, \emph{977}, 173\relax
\mciteBstWouldAddEndPuncttrue
\mciteSetBstMidEndSepPunct{\mcitedefaultmidpunct}
{\mcitedefaultendpunct}{\mcitedefaultseppunct}\relax
\EndOfBibitem
\bibitem[{Perotti} \latin{et~al.}(2026){Perotti}, {Kurtovic}, {Henning}, {Olofsson}, {Arabhavi}, {Schwarz}, {Kanwar}, {van Boekel}, {Kamp}, {Pascucci}, {van Dishoeck}, {G{\"u}del}, {Lagage}, {Barrado}, {Caratti o Garatti}, {Glauser}, {Lahuis}, {Christiaens}, {Franceschi}, {Gasman}, {Grant}, {Jang}, {Kaeufer}, {Morales-Calder{\'o}n}, {Temmink}, and {Vlasblom}]{perotti2026}
{Perotti},~G. \latin{et~al.}  {MINDS. Anatomy of a Water-rich, Inclined, Brown Dwarf Disk: Lack of Abundant Hydrocarbons}. \emph{\apj} \textbf{2026}, \emph{997}, 281\relax
\mciteBstWouldAddEndPuncttrue
\mciteSetBstMidEndSepPunct{\mcitedefaultmidpunct}
{\mcitedefaultendpunct}{\mcitedefaultseppunct}\relax
\EndOfBibitem
\bibitem[{Krijt} \latin{et~al.}(2023){Krijt}, {Kama}, {McClure}, {Teske}, {Bergin}, {Shorttle}, {Walsh}, and {Raymond}]{Krijt23}
{Krijt},~S.; {Kama},~M.; {McClure},~M.; {Teske},~J.; {Bergin},~E.~A.; {Shorttle},~O.; {Walsh},~K.~J.; {Raymond},~S.~N. {Chemical Habitability: Supply and Retention of Life's Essential Elements During Planet Formation}. Protostars and Planets VII. 2023; p 1031\relax
\mciteBstWouldAddEndPuncttrue
\mciteSetBstMidEndSepPunct{\mcitedefaultmidpunct}
{\mcitedefaultendpunct}{\mcitedefaultseppunct}\relax
\EndOfBibitem
\bibitem[{Salyk} \latin{et~al.}(2025){Salyk}, {Pontoppidan}, {Banzatti}, {Bergin}, {Arulanantham}, {Najita}, {Blake}, {Carr}, {Zhang}, and {Xie}]{salyk2025}
{Salyk},~C.; {Pontoppidan},~K.~M.; {Banzatti},~A.; {Bergin},~E.; {Arulanantham},~N.; {Najita},~J.; {Blake},~G.~A.; {Carr},~J.; {Zhang},~K.; {Xie},~C. {Emission from Multiple Molecular Isotopologues in a High-inclination Protoplanetary Disk}. \emph{\aj} \textbf{2025}, \emph{169}, 184\relax
\mciteBstWouldAddEndPuncttrue
\mciteSetBstMidEndSepPunct{\mcitedefaultmidpunct}
{\mcitedefaultendpunct}{\mcitedefaultseppunct}\relax
\EndOfBibitem
\bibitem[{Bergner} \latin{et~al.}(2021){Bergner}, {{\"O}berg}, {Guzm{\'a}n}, {Law}, {Loomis}, {Cataldi}, {Bosman}, {Aikawa}, {Andrews}, {Bergin}, {Booth}, {Cleeves}, {Czekala}, {Huang}, {Ilee}, {Le Gal}, {Long}, {Nomura}, {M{\'e}nard}, {Qi}, {Schwarz}, {Teague}, {Tsukagoshi}, {Walsh}, {Wilner}, and {Yamato}]{Bergner21}
{Bergner},~J.~B. \latin{et~al.}  {Molecules with ALMA at Planet-forming Scales (MAPS). XI. CN and HCN as Tracers of Photochemistry in Disks}. \emph{\apjs} \textbf{2021}, \emph{257}, 11\relax
\mciteBstWouldAddEndPuncttrue
\mciteSetBstMidEndSepPunct{\mcitedefaultmidpunct}
{\mcitedefaultendpunct}{\mcitedefaultseppunct}\relax
\EndOfBibitem
\bibitem[{Salinas} \latin{et~al.}(2016){Salinas}, {Hogerheijde}, {Bergin}, {Cleeves}, {Brinch}, {Blake}, {Lis}, {Melnick}, {Pani{\'c}}, {Pearson}, {Kristensen}, {Y{\i}ld{\i}z}, and {van Dishoeck}]{salinas2016}
{Salinas},~V.~N.; {Hogerheijde},~M.~R.; {Bergin},~E.~A.; {Cleeves},~L.~I.; {Brinch},~C.; {Blake},~G.~A.; {Lis},~D.~C.; {Melnick},~G.~J.; {Pani{\'c}},~O.; {Pearson},~J.~C.; {Kristensen},~L.; {Y{\i}ld{\i}z},~U.~A.; {van Dishoeck},~E.~F. {First detection of gas-phase ammonia in a planet-forming disk. NH$_{3}$, N$_{2}$H$^{+}$, and H$_{2}$O in the disk around TW Hydrae}. \emph{\aap} \textbf{2016}, \emph{591}, A122\relax
\mciteBstWouldAddEndPuncttrue
\mciteSetBstMidEndSepPunct{\mcitedefaultmidpunct}
{\mcitedefaultendpunct}{\mcitedefaultseppunct}\relax
\EndOfBibitem
\bibitem[{Schwarz} and {Bergin}(2014){Schwarz}, and {Bergin}]{Schwarz14}
{Schwarz},~K.~R.; {Bergin},~E.~A. {The Effects of Initial Abundances on Nitrogen in Protoplanetary Disks}. \emph{\apj} \textbf{2014}, \emph{797}, 113\relax
\mciteBstWouldAddEndPuncttrue
\mciteSetBstMidEndSepPunct{\mcitedefaultmidpunct}
{\mcitedefaultendpunct}{\mcitedefaultseppunct}\relax
\EndOfBibitem
\bibitem[{Kanwar} \latin{et~al.}(2025){Kanwar}, {Woitke}, {Kamp}, {Rimmer}, and {Helling}]{kanwar2025}
{Kanwar},~J.; {Woitke},~P.; {Kamp},~I.; {Rimmer},~P.; {Helling},~C. {Can thermodynamic equilibrium be established in planet-forming disks?} \emph{\aap} \textbf{2025}, \emph{698}, A294\relax
\mciteBstWouldAddEndPuncttrue
\mciteSetBstMidEndSepPunct{\mcitedefaultmidpunct}
{\mcitedefaultendpunct}{\mcitedefaultseppunct}\relax
\EndOfBibitem
\bibitem[{van Gelder} \latin{et~al.}(2024){van Gelder}, {Francis}, {van Dishoeck}, {Tychoniec}, {Ray}, {Beuther}, {Caratti o Garatti}, {Chen}, {Devaraj}, {Gieser}, {Justtanont}, {Kavanagh}, {Nazari}, {Reyes}, {Rocha}, {Slavicinska}, {G{\"u}del}, {Henning}, {Lagage}, and {Wright}]{vangelder2024}
{van Gelder},~M.~L. \latin{et~al.}  {JWST Observations of Young protoStars (JOYS): Overview of gaseous molecular emission and absorption in low-mass protostars}. \emph{\aap} \textbf{2024}, \emph{692}, A197\relax
\mciteBstWouldAddEndPuncttrue
\mciteSetBstMidEndSepPunct{\mcitedefaultmidpunct}
{\mcitedefaultendpunct}{\mcitedefaultseppunct}\relax
\EndOfBibitem
\bibitem[{Najita} \latin{et~al.}(2021){Najita}, {Carr}, {Brittain}, {Lacy}, {Richter}, and {Doppmann}]{najita2021}
{Najita},~J.~R.; {Carr},~J.~S.; {Brittain},~S.~D.; {Lacy},~J.~H.; {Richter},~M.~J.; {Doppmann},~G.~W. {High-resolution Mid-infrared Spectroscopy of GV Tau N: Surface Accretion and Detection of NH$_{3}$ in a Young Protoplanetary Disk}. \emph{\apj} \textbf{2021}, \emph{908}, 171\relax
\mciteBstWouldAddEndPuncttrue
\mciteSetBstMidEndSepPunct{\mcitedefaultmidpunct}
{\mcitedefaultendpunct}{\mcitedefaultseppunct}\relax
\EndOfBibitem
\bibitem[{Kaeufer} \latin{et~al.}(2024){Kaeufer}, {Woitke}, {Kamp}, {Kanwar}, and {Min}]{kaeufer2024}
{Kaeufer},~T.; {Woitke},~P.; {Kamp},~I.; {Kanwar},~J.; {Min},~M. {Disentangling the dust and gas contributions of the JWST/MIRI spectrum of Sz 28}. \emph{\aap} \textbf{2024}, \emph{690}, A100\relax
\mciteBstWouldAddEndPuncttrue
\mciteSetBstMidEndSepPunct{\mcitedefaultmidpunct}
{\mcitedefaultendpunct}{\mcitedefaultseppunct}\relax
\EndOfBibitem
\bibitem[{Pontoppidan} \latin{et~al.}(2019){Pontoppidan}, {Salyk}, {Banzatti}, {Blake}, {Walsh}, {Lacy}, and {Richter}]{pontoppidan2019}
{Pontoppidan},~K.~M.; {Salyk},~C.; {Banzatti},~A.; {Blake},~G.~A.; {Walsh},~C.; {Lacy},~J.~H.; {Richter},~M.~J. {The Nitrogen Carrier in Inner Protoplanetary Disks}. \emph{\apj} \textbf{2019}, \emph{874}, 92\relax
\mciteBstWouldAddEndPuncttrue
\mciteSetBstMidEndSepPunct{\mcitedefaultmidpunct}
{\mcitedefaultendpunct}{\mcitedefaultseppunct}\relax
\EndOfBibitem
\bibitem[{Arabhavi} \latin{et~al.}(2026){Arabhavi}, {Kamp}, {van Dishoeck}, {Woitke}, {Rab}, {Thi}, {Kaeufer}, {Kanwar}, {Tabone}, {Esteve}, and {Vlasblom}]{arabhavi2026}
{Arabhavi},~A.~M.; {Kamp},~I.; {van Dishoeck},~E.~F.; {Woitke},~P.; {Rab},~C.; {Thi},~W.-F.; {Kaeufer},~T.; {Kanwar},~J.; {Tabone},~B.; {Esteve},~P.; {Vlasblom},~M. {Molecular diagnostics for the mid-infrared emission of planet-forming disks. Carbon and oxygen elemental abundances}. \emph{arXiv e-prints} \textbf{2026}, arXiv:2602.16030\relax
\mciteBstWouldAddEndPuncttrue
\mciteSetBstMidEndSepPunct{\mcitedefaultmidpunct}
{\mcitedefaultendpunct}{\mcitedefaultseppunct}\relax
\EndOfBibitem
\bibitem[{Kalyaan} \latin{et~al.}(2021){Kalyaan}, {Pinilla}, {Krijt}, {Mulders}, and {Banzatti}]{kalyaan2021}
{Kalyaan},~A.; {Pinilla},~P.; {Krijt},~S.; {Mulders},~G.~D.; {Banzatti},~A. {Linking Outer Disk Pebble Dynamics and Gaps to Inner Disk Water Enrichment}. \emph{\apj} \textbf{2021}, \emph{921}, 84\relax
\mciteBstWouldAddEndPuncttrue
\mciteSetBstMidEndSepPunct{\mcitedefaultmidpunct}
{\mcitedefaultendpunct}{\mcitedefaultseppunct}\relax
\EndOfBibitem
\bibitem[{Krijt} \latin{et~al.}(2025){Krijt}, {Banzatti}, {Zhang}, {Pinilla}, {Kaeufer}, {Bergin}, {Salyk}, {Pontoppidan}, {Blake}, {Long}, {Huang}, {Colmenares}, {Williams}, {Houge}, {Narang}, {Vioque}, {Lambrechts}, {Cleeves}, {{\"O}berg}, and {The Jdiscs Collaboration}]{krijt2025}
{Krijt},~S. \latin{et~al.}  {Cosmic Cascades: How Disk Substructure Regulates the Flow of Water to Inner Planetary Systems}. \emph{\apjl} \textbf{2025}, \emph{990}, L72\relax
\mciteBstWouldAddEndPuncttrue
\mciteSetBstMidEndSepPunct{\mcitedefaultmidpunct}
{\mcitedefaultendpunct}{\mcitedefaultseppunct}\relax
\EndOfBibitem
\bibitem[{Pinilla} \latin{et~al.}(2012){Pinilla}, {Birnstiel}, {Ricci}, {Dullemond}, {Uribe}, {Testi}, and {Natta}]{Pinilla12}
{Pinilla},~P.; {Birnstiel},~T.; {Ricci},~L.; {Dullemond},~C.~P.; {Uribe},~A.~L.; {Testi},~L.; {Natta},~A. {Trapping dust particles in the outer regions of protoplanetary disks}. \emph{\aap} \textbf{2012}, \emph{538}, A114\relax
\mciteBstWouldAddEndPuncttrue
\mciteSetBstMidEndSepPunct{\mcitedefaultmidpunct}
{\mcitedefaultendpunct}{\mcitedefaultseppunct}\relax
\EndOfBibitem
\bibitem[{Andrews}(2020)]{Andrews20}
{Andrews},~S.~M. {Observations of Protoplanetary Disk Structures}. \emph{\araa} \textbf{2020}, \emph{58}, 483--528\relax
\mciteBstWouldAddEndPuncttrue
\mciteSetBstMidEndSepPunct{\mcitedefaultmidpunct}
{\mcitedefaultendpunct}{\mcitedefaultseppunct}\relax
\EndOfBibitem
\bibitem[{Booth} and {Ilee}(2019){Booth}, and {Ilee}]{Booth19}
{Booth},~R.~A.; {Ilee},~J.~D. {Planet-forming material in a protoplanetary disc: the interplay between chemical evolution and pebble drift}. \emph{\mnras} \textbf{2019}, \emph{487}, 3998--4011\relax
\mciteBstWouldAddEndPuncttrue
\mciteSetBstMidEndSepPunct{\mcitedefaultmidpunct}
{\mcitedefaultendpunct}{\mcitedefaultseppunct}\relax
\EndOfBibitem
\bibitem[{Henning} \latin{et~al.}(2024){Henning}, {Kamp}, {Samland}, {Arabhavi}, {Kanwar}, {van Dishoeck}, {G{\"u}del}, {Lagage}, {Waelkens}, {Abergel}, {Absil}, {Barrado}, {Boccaletti}, {Bouwman}, {Caratti o Garatti}, {Geers}, {Glauser}, {Lahuis}, {Mueller}, {Nehm{\'e}}, {Olofsson}, {Pantin}, {Ray}, {Scheithauer}, {Vandenbussche}, {Waters}, {Wright}, {Argyriou}, {Christiaens}, {Franceschi}, {Gasman}, {Grant}, {Guadarrama}, {Jang}, {Morales-Calder{\'o}n}, {Pawellek}, {Perotti}, {Rodgers-Lee}, {Schreiber}, {Schwarz}, {Tabone}, {Temmink}, {Vlasblom}, {Colina}, {Greve}, and {{\"O}stlin}]{henning2024}
{Henning},~T. \latin{et~al.}  {MINDS: The JWST MIRI Mid-INfrared Disk Survey}. \emph{\pasp} \textbf{2024}, \emph{136}, 054302\relax
\mciteBstWouldAddEndPuncttrue
\mciteSetBstMidEndSepPunct{\mcitedefaultmidpunct}
{\mcitedefaultendpunct}{\mcitedefaultseppunct}\relax
\EndOfBibitem
\bibitem[{Bottinelli} \latin{et~al.}(2010){Bottinelli}, {Boogert}, {Bouwman}, {Beckwith}, {van Dishoeck}, {{\"O}berg}, {Pontoppidan}, {Linnartz}, {Blake}, {Evans}, and {Lahuis}]{Bottinelli10}
{Bottinelli},~S.; {Boogert},~A.~C.~A.; {Bouwman},~J.; {Beckwith},~M.; {van Dishoeck},~E.~F.; {{\"O}berg},~K.~I.; {Pontoppidan},~K.~M.; {Linnartz},~H.; {Blake},~G.~A.; {Evans},~N.~J.,~II; {Lahuis},~F. {The c2d Spitzer Spectroscopic Survey of Ices Around Low-mass Young Stellar Objects. IV. NH$_{3}$ and CH$_{3}$OH}. \emph{\apj} \textbf{2010}, \emph{718}, 1100--1117\relax
\mciteBstWouldAddEndPuncttrue
\mciteSetBstMidEndSepPunct{\mcitedefaultmidpunct}
{\mcitedefaultendpunct}{\mcitedefaultseppunct}\relax
\EndOfBibitem
\bibitem[{Boogert} \latin{et~al.}(2015){Boogert}, {Gerakines}, and {Whittet}]{boogert2015}
{Boogert},~A.~C.~A.; {Gerakines},~P.~A.; {Whittet},~D. C.~B. {Observations of the icy universe.} \emph{\araa} \textbf{2015}, \emph{53}, 541--581\relax
\mciteBstWouldAddEndPuncttrue
\mciteSetBstMidEndSepPunct{\mcitedefaultmidpunct}
{\mcitedefaultendpunct}{\mcitedefaultseppunct}\relax
\EndOfBibitem
\bibitem[{Altwegg} \latin{et~al.}(2019){Altwegg}, {Balsiger}, and {Fuselier}]{Altwegg19}
{Altwegg},~K.; {Balsiger},~H.; {Fuselier},~S.~A. {Cometary Chemistry and the Origin of Icy Solar System Bodies: The View After Rosetta}. \emph{\araa} \textbf{2019}, \emph{57}, 113--155\relax
\mciteBstWouldAddEndPuncttrue
\mciteSetBstMidEndSepPunct{\mcitedefaultmidpunct}
{\mcitedefaultendpunct}{\mcitedefaultseppunct}\relax
\EndOfBibitem
\bibitem[{Altwegg} \latin{et~al.}(2022){Altwegg}, {Combi}, {Fuselier}, {H{\"a}nni}, {De Keyser}, {Mahjoub}, {M{\"u}ller}, {Pestoni}, {Rubin}, and {Wampfler}]{Altwegg22}
{Altwegg},~K.; {Combi},~M.; {Fuselier},~S.~A.; {H{\"a}nni},~N.; {De Keyser},~J.; {Mahjoub},~A.; {M{\"u}ller},~D.~R.; {Pestoni},~B.; {Rubin},~M.; {Wampfler},~S.~F. {Abundant ammonium hydrosulphide embedded in cometary dust grains}. \emph{\mnras} \textbf{2022}, \emph{516}, 3900--3910\relax
\mciteBstWouldAddEndPuncttrue
\mciteSetBstMidEndSepPunct{\mcitedefaultmidpunct}
{\mcitedefaultendpunct}{\mcitedefaultseppunct}\relax
\EndOfBibitem
\bibitem[{Bosman} \latin{et~al.}(2017){Bosman}, {Bruderer}, and {van Dishoeck}]{bosman2017}
{Bosman},~A.~D.; {Bruderer},~S.; {van Dishoeck},~E.~F. {CO$_{2}$ infrared emission as a diagnostic of planet-forming regions of disks}. \emph{\aap} \textbf{2017}, \emph{601}, A36\relax
\mciteBstWouldAddEndPuncttrue
\mciteSetBstMidEndSepPunct{\mcitedefaultmidpunct}
{\mcitedefaultendpunct}{\mcitedefaultseppunct}\relax
\EndOfBibitem
\bibitem[{Vlasblom} \latin{et~al.}(2025){Vlasblom}, {Temmink}, {Grant}, {Kurtovic}, {Sellek}, {van Dishoeck}, {G{\"u}del}, {Henning}, {Lagage}, {Barrado}, {Caratti o Garatti}, {Glauser}, {Kamp}, {Lahuis}, {Olofsson}, {Arabhavi}, {Christiaens}, {Gasman}, {Jang}, {Morales-Calder{\'o}n}, {Perotti}, {Schwarz}, and {Tabone}]{vlasblom2025_CXTau}
{Vlasblom},~M. \latin{et~al.}  {MINDS. JWST-MIRI reveals a peculiar CO$_{2}$-rich chemistry in the drift-dominated disk CX Tau}. \emph{\aap} \textbf{2025}, \emph{693}, A278\relax
\mciteBstWouldAddEndPuncttrue
\mciteSetBstMidEndSepPunct{\mcitedefaultmidpunct}
{\mcitedefaultendpunct}{\mcitedefaultseppunct}\relax
\EndOfBibitem
\bibitem[{Sellek} \latin{et~al.}(2025){Sellek}, {Vlasblom}, and {van Dishoeck}]{sellek2024}
{Sellek},~A.~D.; {Vlasblom},~M.; {van Dishoeck},~E.~F. {CO$_{2}$-rich protoplanetary discs as a probe of dust radial drift and trapping}. \emph{\aap} \textbf{2025}, \emph{694}, A79\relax
\mciteBstWouldAddEndPuncttrue
\mciteSetBstMidEndSepPunct{\mcitedefaultmidpunct}
{\mcitedefaultendpunct}{\mcitedefaultseppunct}\relax
\EndOfBibitem
\bibitem[{Ag{\'u}ndez} \latin{et~al.}(2008){Ag{\'u}ndez}, {Cernicharo}, and {Goicoechea}]{Agundez08}
{Ag{\'u}ndez},~M.; {Cernicharo},~J.; {Goicoechea},~J.~R. {Formation of simple organic molecules in inner T Tauri disks}. \emph{\aap} \textbf{2008}, \emph{483}, 831--837\relax
\mciteBstWouldAddEndPuncttrue
\mciteSetBstMidEndSepPunct{\mcitedefaultmidpunct}
{\mcitedefaultendpunct}{\mcitedefaultseppunct}\relax
\EndOfBibitem
\bibitem[{Woitke} \latin{et~al.}(2016){Woitke}, {Min}, {Pinte}, {Thi}, {Kamp}, {Rab}, {Anthonioz}, {Antonellini}, {Baldovin-Saavedra}, {Carmona}, {Dominik}, {Dionatos}, {Greaves}, {G{\"u}del}, {Ilee}, {Liebhart}, {M{\'e}nard}, {Rigon}, {Waters}, {Aresu}, {Meijerink}, and {Spaans}]{woitke2016}
{Woitke},~P. \latin{et~al.}  {Consistent dust and gas models for protoplanetary disks. I. Disk shape, dust settling, opacities, and PAHs}. \emph{\aap} \textbf{2016}, \emph{586}, A103\relax
\mciteBstWouldAddEndPuncttrue
\mciteSetBstMidEndSepPunct{\mcitedefaultmidpunct}
{\mcitedefaultendpunct}{\mcitedefaultseppunct}\relax
\EndOfBibitem
\bibitem[{Kamp} \latin{et~al.}(2017){Kamp}, {Thi}, {Woitke}, {Rab}, {Bouma}, and {M{\'e}nard}]{Kamp17}
{Kamp},~I.; {Thi},~W.-F.; {Woitke},~P.; {Rab},~C.; {Bouma},~S.; {M{\'e}nard},~F. {Consistent dust and gas models for protoplanetary disks. II. Chemical networks and rates}. \emph{\aap} \textbf{2017}, \emph{607}, A41\relax
\mciteBstWouldAddEndPuncttrue
\mciteSetBstMidEndSepPunct{\mcitedefaultmidpunct}
{\mcitedefaultendpunct}{\mcitedefaultseppunct}\relax
\EndOfBibitem
\bibitem[{Woitke} \latin{et~al.}(2009){Woitke}, {Kamp}, and {Thi}]{woitke2009}
{Woitke},~P.; {Kamp},~I.; {Thi},~W.~F. {Radiation thermo-chemical models of protoplanetary disks. I. Hydrostatic disk structure and inner rim}. \emph{\aap} \textbf{2009}, \emph{501}, 383--406\relax
\mciteBstWouldAddEndPuncttrue
\mciteSetBstMidEndSepPunct{\mcitedefaultmidpunct}
{\mcitedefaultendpunct}{\mcitedefaultseppunct}\relax
\EndOfBibitem
\bibitem[{Manara} \latin{et~al.}(2023){Manara}, {Ansdell}, {Rosotti}, {Hughes}, {Armitage}, {Lodato}, and {Williams}]{manara2023}
{Manara},~C.~F.; {Ansdell},~M.; {Rosotti},~G.~P.; {Hughes},~A.~M.; {Armitage},~P.~J.; {Lodato},~G.; {Williams},~J.~P. {Demographics of Young Stars and their Protoplanetary Disks: Lessons Learned on Disk Evolution and its Connection to Planet Formation}. Protostars and Planets VII. 2023; p 539\relax
\mciteBstWouldAddEndPuncttrue
\mciteSetBstMidEndSepPunct{\mcitedefaultmidpunct}
{\mcitedefaultendpunct}{\mcitedefaultseppunct}\relax
\EndOfBibitem
\bibitem[{Woitke} \latin{et~al.}(2024){Woitke}, {Thi}, {Arabhavi}, {Kamp}, {K{\'o}sp{\'a}l}, and {{\'A}brah{\'a}m}]{woitke2024}
{Woitke},~P.; {Thi},~W.~F.; {Arabhavi},~A.~M.; {Kamp},~I.; {K{\'o}sp{\'a}l},~{\'A}.; {{\'A}brah{\'a}m},~P. {2D disc modelling of the JWST line spectrum of EX Lupi}. \emph{\aap} \textbf{2024}, \emph{683}, A219\relax
\mciteBstWouldAddEndPuncttrue
\mciteSetBstMidEndSepPunct{\mcitedefaultmidpunct}
{\mcitedefaultendpunct}{\mcitedefaultseppunct}\relax
\EndOfBibitem
\bibitem[{Meijerink} \latin{et~al.}(2009){Meijerink}, {Pontoppidan}, {Blake}, {Poelman}, and {Dullemond}]{meijerink2009}
{Meijerink},~R.; {Pontoppidan},~K.~M.; {Blake},~G.~A.; {Poelman},~D.~R.; {Dullemond},~C.~P. {Radiative Transfer Models of Mid-Infrared H$_{2}$O Lines in the Planet-Forming Region of Circumstellar Disks}. \emph{\apj} \textbf{2009}, \emph{704}, 1471--1481\relax
\mciteBstWouldAddEndPuncttrue
\mciteSetBstMidEndSepPunct{\mcitedefaultmidpunct}
{\mcitedefaultendpunct}{\mcitedefaultseppunct}\relax
\EndOfBibitem
\bibitem[{Riols} and {Lesur}(2018){Riols}, and {Lesur}]{riols_lesur_2018}
{Riols},~A.; {Lesur},~G. {Dust settling and rings in the outer regions of protoplanetary discs subject to ambipolar diffusion}. \emph{\aap} \textbf{2018}, \emph{617}, A117\relax
\mciteBstWouldAddEndPuncttrue
\mciteSetBstMidEndSepPunct{\mcitedefaultmidpunct}
{\mcitedefaultendpunct}{\mcitedefaultseppunct}\relax
\EndOfBibitem
\bibitem[{Kanwar} \latin{et~al.}(2024){Kanwar}, {Kamp}, {Woitke}, {Rab}, {Thi}, and {Min}]{kanwar2024_model}
{Kanwar},~J.; {Kamp},~I.; {Woitke},~P.; {Rab},~C.; {Thi},~W.~F.; {Min},~M. {Hydrocarbon chemistry in the inner regions of planet-forming disks}. \emph{\aap} \textbf{2024}, \emph{681}, A22\relax
\mciteBstWouldAddEndPuncttrue
\mciteSetBstMidEndSepPunct{\mcitedefaultmidpunct}
{\mcitedefaultendpunct}{\mcitedefaultseppunct}\relax
\EndOfBibitem
\bibitem[{McElroy} \latin{et~al.}(2013){McElroy}, {Walsh}, {Markwick}, {Cordiner}, {Smith}, and {Millar}]{mcelroy2013_UMIST12}
{McElroy},~D.; {Walsh},~C.; {Markwick},~A.~J.; {Cordiner},~M.~A.; {Smith},~K.; {Millar},~T.~J. {The UMIST database for astrochemistry 2012}. \emph{\aap} \textbf{2013}, \emph{550}, A36\relax
\mciteBstWouldAddEndPuncttrue
\mciteSetBstMidEndSepPunct{\mcitedefaultmidpunct}
{\mcitedefaultendpunct}{\mcitedefaultseppunct}\relax
\EndOfBibitem
\bibitem[{Heays} \latin{et~al.}(2017){Heays}, {Bosman}, and {van Dishoeck}]{heays2017}
{Heays},~A.~N.; {Bosman},~A.~D.; {van Dishoeck},~E.~F. {Photodissociation and photoionisation of atoms and molecules of astrophysical interest}. \emph{\aap} \textbf{2017}, \emph{602}, A105\relax
\mciteBstWouldAddEndPuncttrue
\mciteSetBstMidEndSepPunct{\mcitedefaultmidpunct}
{\mcitedefaultendpunct}{\mcitedefaultseppunct}\relax
\EndOfBibitem
\bibitem[{Li} \latin{et~al.}(2013){Li}, {Heays}, {Visser}, {Ubachs}, {Lewis}, {Gibson}, and {van Dishoeck}]{li2013}
{Li},~X.; {Heays},~A.~N.; {Visser},~R.; {Ubachs},~W.; {Lewis},~B.~R.; {Gibson},~S.~T.; {van Dishoeck},~E.~F. {Photodissociation of interstellar N$_{2}$}. \emph{\aap} \textbf{2013}, \emph{555}, A14\relax
\mciteBstWouldAddEndPuncttrue
\mciteSetBstMidEndSepPunct{\mcitedefaultmidpunct}
{\mcitedefaultendpunct}{\mcitedefaultseppunct}\relax
\EndOfBibitem
\bibitem[{Davidson} and {Henson}(1990){Davidson}, and {Henson}]{davidson1990}
{Davidson},~D.~F.; {Henson},~R.~K. {High temperature reaction rate coefficients derived from N-atom ARAS measurements and excimer photolysis of NO}. \emph{Int. J. Chem. Kinet.} \textbf{1990}, \emph{22}, 843\relax
\mciteBstWouldAddEndPuncttrue
\mciteSetBstMidEndSepPunct{\mcitedefaultmidpunct}
{\mcitedefaultendpunct}{\mcitedefaultseppunct}\relax
\EndOfBibitem
\bibitem[{Tielens} and {Hollenbach}(1985){Tielens}, and {Hollenbach}]{tielens1985}
{Tielens},~A.~G.~G.~M.; {Hollenbach},~D. {Photodissociation regions. I. Basic model.} \emph{\apj} \textbf{1985}, \emph{291}, 722--746\relax
\mciteBstWouldAddEndPuncttrue
\mciteSetBstMidEndSepPunct{\mcitedefaultmidpunct}
{\mcitedefaultendpunct}{\mcitedefaultseppunct}\relax
\EndOfBibitem
\bibitem[{Gordon} \latin{et~al.}(2022){Gordon}, {Rothman}, {Hargreaves}, {Hashemi}, {Karlovets}, {Skinner}, {Conway}, {Hill}, {Kochanov}, {Tan}, {Wcis{\l}o}, {Finenko}, {Nelson}, {Bernath}, {Birk}, {Boudon}, {Campargue}, {Chance}, {Coustenis}, {Drouin}, {Flaud}, {Gamache}, {Hodges}, {Jacquemart}, {Mlawer}, {Nikitin}, {Perevalov}, {Rotger}, {Tennyson}, {Toon}, {Tran}, {Tyuterev}, {Adkins}, {Baker}, {Barbe}, {Can{\`e}}, {Cs{\'a}sz{\'a}r}, {Dudaryonok}, {Egorov}, {Fleisher}, {Fleurbaey}, {Foltynowicz}, {Furtenbacher}, {Harrison}, {Hartmann}, {Horneman}, {Huang}, {Karman}, {Karns}, {Kassi}, {Kleiner}, {Kofman}, {Kwabia-Tchana}, {Lavrentieva}, {Lee}, {Long}, {Lukashevskaya}, {Lyulin}, {Makhnev}, {Matt}, {Massie}, {Melosso}, {Mikhailenko}, {Mondelain}, {M{\"u}ller}, {Naumenko}, {Perrin}, {Polyansky}, {Raddaoui}, {Raston}, {Reed}, {Rey}, {Richard}, {T{\'o}bi{\'a}s}, {Sadiek}, {Schwenke}, {Starikova}, {Sung}, {Tamassia}, {Tashkun}, {Vander Auwera}, {Vasilenko}, {Vigasin}, {Villanueva}, {Vispoel}, {Wagner},
  {Yachmenev}, and {Yurchenko}]{2022JQSRT.27707949G}
{Gordon},~I.~E. \latin{et~al.}  {The HITRAN2020 molecular spectroscopic database}. \emph{\jqsrt} \textbf{2022}, \emph{277}, 107949\relax
\mciteBstWouldAddEndPuncttrue
\mciteSetBstMidEndSepPunct{\mcitedefaultmidpunct}
{\mcitedefaultendpunct}{\mcitedefaultseppunct}\relax
\EndOfBibitem
\bibitem[{Grant} \latin{et~al.}(2023){Grant}, {van Dishoeck}, {Tabone}, {Gasman}, {Henning}, {Kamp}, {G{\"u}del}, {Lagage}, {Bettoni}, {Perotti}, {Christiaens}, {Samland}, {Arabhavi}, {Argyriou}, {Abergel}, {Absil}, {Barrado}, {Boccaletti}, {Bouwman}, {o Garatti}, {Geers}, {Glauser}, {Guadarrama}, {Jang}, {Kanwar}, {Lahuis}, {Morales-Calder{\'o}n}, {Mueller}, {Nehm{\'e}}, {Olofsson}, {Pantin}, {Pawellek}, {Ray}, {Rodgers-Lee}, {Scheithauer}, {Schreiber}, {Schwarz}, {Temmink}, {Vandenbussche}, {Vlasblom}, {Waters}, {Wright}, {Colina}, {Greve}, {Justannont}, and {{\"O}stlin}]{grant2023}
{Grant},~S.~L. \latin{et~al.}  {MINDS. The Detection of $^{13}$CO$_{2}$ with JWST-MIRI Indicates Abundant CO$_{2}$ in a Protoplanetary Disk}. \emph{\apjl} \textbf{2023}, \emph{947}, L6\relax
\mciteBstWouldAddEndPuncttrue
\mciteSetBstMidEndSepPunct{\mcitedefaultmidpunct}
{\mcitedefaultendpunct}{\mcitedefaultseppunct}\relax
\EndOfBibitem
\bibitem[{Gasman} \latin{et~al.}(2023){Gasman}, {van Dishoeck}, {Grant}, {Temmink}, {Tabone}, {Henning}, {Kamp}, {G{\"u}del}, {Lagage}, {Perotti}, {Christiaens}, {Samland}, {Arabhavi}, {Argyriou}, {Abergel}, {Absil}, {Barrado}, {Boccaletti}, {Bouwman}, {Caratti o Garatti}, {Geers}, {Glauser}, {Guadarrama}, {Jang}, {Kanwar}, {Lahuis}, {Morales-Calder{\'o}n}, {Mueller}, {Nehm{\'e}}, {Olofsson}, {Pantin}, {Pawellek}, {Ray}, {Rodgers-Lee}, {Scheithauer}, {Schreiber}, {Schwarz}, {Vandenbussche}, {Vlasblom}, {Waters}, {Wright}, {Colina}, {Greve}, and {{\"O}stlin}]{gasman2023b}
{Gasman},~D. \latin{et~al.}  {MINDS. Abundant water and varying C/O across the disk of Sz 98 as seen by JWST/MIRI}. \emph{\aap} \textbf{2023}, \emph{679}, A117\relax
\mciteBstWouldAddEndPuncttrue
\mciteSetBstMidEndSepPunct{\mcitedefaultmidpunct}
{\mcitedefaultendpunct}{\mcitedefaultseppunct}\relax
\EndOfBibitem
\bibitem[{Wakelam} \latin{et~al.}(2012){Wakelam}, {Herbst}, {Loison}, {Smith}, {Chandrasekaran}, {Pavone}, {Adams}, {Bacchus-Montabonel}, {Bergeat}, {B{\'e}roff}, {Bierbaum}, {Chabot}, {Dalgarno}, {van Dishoeck}, {Faure}, {Geppert}, {Gerlich}, {Galli}, {H{\'e}brard}, {Hersant}, {Hickson}, {Honvault}, {Klippenstein}, {Le Picard}, {Nyman}, {Pernot}, {Schlemmer}, {Selsis}, {Sims}, {Talbi}, {Tennyson}, {Troe}, {Wester}, and {Wiesenfeld}]{wakelam2012_KIDA}
{Wakelam},~V. \latin{et~al.}  {A KInetic Database for Astrochemistry (KIDA)}. \emph{\apjs} \textbf{2012}, \emph{199}, 21\relax
\mciteBstWouldAddEndPuncttrue
\mciteSetBstMidEndSepPunct{\mcitedefaultmidpunct}
{\mcitedefaultendpunct}{\mcitedefaultseppunct}\relax
\EndOfBibitem
\bibitem[{Charnley}(1997)]{charnley1997}
{Charnley},~S.~B. {Sulfuretted Molecules in Hot Cores}. \emph{\apj} \textbf{1997}, \emph{481}, 396--405\relax
\mciteBstWouldAddEndPuncttrue
\mciteSetBstMidEndSepPunct{\mcitedefaultmidpunct}
{\mcitedefaultendpunct}{\mcitedefaultseppunct}\relax
\EndOfBibitem
\bibitem[{van Dishoeck} \latin{et~al.}(2013){van Dishoeck}, {Herbst}, and {Neufeld}]{vandishoeck2013}
{van Dishoeck},~E.~F.; {Herbst},~E.; {Neufeld},~D.~A. {Interstellar Water Chemistry: From Laboratory to Observations}. \emph{Chemical Reviews} \textbf{2013}, \emph{113}, 9043--9085\relax
\mciteBstWouldAddEndPuncttrue
\mciteSetBstMidEndSepPunct{\mcitedefaultmidpunct}
{\mcitedefaultendpunct}{\mcitedefaultseppunct}\relax
\EndOfBibitem
\bibitem[{Bast} \latin{et~al.}(2013){Bast}, {Lahuis}, {van Dishoeck}, and {Tielens}]{bast2013}
{Bast},~J.~E.; {Lahuis},~F.; {van Dishoeck},~E.~F.; {Tielens},~A.~G.~G.~M. {Exploring organic chemistry in planet-forming zones}. \emph{\aap} \textbf{2013}, \emph{551}, A118\relax
\mciteBstWouldAddEndPuncttrue
\mciteSetBstMidEndSepPunct{\mcitedefaultmidpunct}
{\mcitedefaultendpunct}{\mcitedefaultseppunct}\relax
\EndOfBibitem
\bibitem[{Baulch} \latin{et~al.}(1994){Baulch}, {Cobos}, {Cox}, {Frank}, {Hayman}, {Just}, {Kerr}, {Murrells}, {Pilling}, {Troe}, {Walker}, and {Warnatz}]{baulch1994}
{Baulch},~D.~L.; {Cobos},~C.~J.; {Cox},~R.~A.; {Frank},~P.; {Hayman},~G.; {Just},~T.; {Kerr},~J.~A.; {Murrells},~T.; {Pilling},~M.~J.; {Troe},~J.; {Walker},~R.~W.; {Warnatz},~J. {Evaluated Kinetic Data for Combustion Modeling. Supplement I}. \emph{J. Phys. Chem. Ref. Data} \textbf{1994}, \emph{23}, 847--848\relax
\mciteBstWouldAddEndPuncttrue
\mciteSetBstMidEndSepPunct{\mcitedefaultmidpunct}
{\mcitedefaultendpunct}{\mcitedefaultseppunct}\relax
\EndOfBibitem
\bibitem[{Kanwar} \latin{et~al.}(2026){Kanwar}, {Kamp}, {Woitke}, {van Dishoeck}, {Henning}, {Liu}, {Kaeufer}, {Tabone}, {G{\"u}del}, {Barrado}, {Arabhavi}, {Franceschi}, and {Vlasblom}]{kanwar2026}
{Kanwar},~J.; {Kamp},~I.; {Woitke},~P.; {van Dishoeck},~E.~F.; {Henning},~T.; {Liu},~Y.; {Kaeufer},~T.; {Tabone},~B.; {G{\"u}del},~M.; {Barrado},~D.; {Arabhavi},~A.~M.; {Franceschi},~R.; {Vlasblom},~M. {MINDS: Strong oxygen depletion in the inner regions of a very low-mass star disk?} \emph{\aap} \textbf{2026}, \emph{705}, A222\relax
\mciteBstWouldAddEndPuncttrue
\mciteSetBstMidEndSepPunct{\mcitedefaultmidpunct}
{\mcitedefaultendpunct}{\mcitedefaultseppunct}\relax
\EndOfBibitem
\bibitem[{Jang} \latin{et~al.}(2025){Jang}, {Arabhavi}, {Kaeufer}, {Waters}, {Kamp}, {Henning}, {Caratti o Garatti}, {van Dishoeck}, {Perotti}, {Kanwar}, {G{\"u}del}, {Morales-Calder{\'o}n}, {Grant}, and {Christiaens}]{jang2025}
{Jang},~H.; {Arabhavi},~A.~M.; {Kaeufer},~T.; {Waters},~R.; {Kamp},~I.; {Henning},~T.; {Caratti o Garatti},~A.; {van Dishoeck},~E.~F.; {Perotti},~G.; {Kanwar},~J.; {G{\"u}del},~M.; {Morales-Calder{\'o}n},~M.; {Grant},~S.~L.; {Christiaens},~V. {MINDS: The very low-mass star and brown dwarf sample: II. Probing disk settling, dust properties, and dust-gas interplay with JWST/MIRI}. \emph{\aap} \textbf{2025}, \emph{703}, A53\relax
\mciteBstWouldAddEndPuncttrue
\mciteSetBstMidEndSepPunct{\mcitedefaultmidpunct}
{\mcitedefaultendpunct}{\mcitedefaultseppunct}\relax
\EndOfBibitem
\bibitem[{Gasman} \latin{et~al.}(2025){Gasman}, {Temmink}, {van Dishoeck}, {Kurtovic}, {Grant}, {Sellek}, {Tabone}, {Henning}, {Kamp}, {G{\"u}del}, {Barrado}, {Caratti o Garatti}, {Glauser}, {Waters}, {Arabhavi}, {Jang}, {Kanwar}, {Lienert}, {Perotti}, {Schwarz}, and {Vlasblom}]{gasman2025}
{Gasman},~D. \latin{et~al.}  {MINDS: The influence of outer dust disc structure on the volatile delivery to the inner disc}. \emph{\aap} \textbf{2025}, \emph{694}, A147\relax
\mciteBstWouldAddEndPuncttrue
\mciteSetBstMidEndSepPunct{\mcitedefaultmidpunct}
{\mcitedefaultendpunct}{\mcitedefaultseppunct}\relax
\EndOfBibitem
\bibitem[{Testi} \latin{et~al.}(2022){Testi}, {Natta}, {Manara}, {de Gregorio Monsalvo}, {Lodato}, {Lopez}, {Muzic}, {Pascucci}, {Sanchis}, {Miranda}, {Scholz}, {De Simone}, and {Williams}]{testi2022}
{Testi},~L.; {Natta},~A.; {Manara},~C.~F.; {de Gregorio Monsalvo},~I.; {Lodato},~G.; {Lopez},~C.; {Muzic},~K.; {Pascucci},~I.; {Sanchis},~E.; {Miranda},~A.~S.; {Scholz},~A.; {De Simone},~M.; {Williams},~J.~P. {The protoplanetary disk population in the {\ensuremath{\rho}}-Ophiuchi region L1688 and the time evolution of Class II YSOs}. \emph{\aap} \textbf{2022}, \emph{663}, A98\relax
\mciteBstWouldAddEndPuncttrue
\mciteSetBstMidEndSepPunct{\mcitedefaultmidpunct}
{\mcitedefaultendpunct}{\mcitedefaultseppunct}\relax
\EndOfBibitem
\bibitem[{Gondoin}(2006)]{gondoin2006}
{Gondoin},~P. {X-ray emission from T Tauri stars in the Lupus 3 star-forming region}. \emph{\aap} \textbf{2006}, \emph{454}, 595--607\relax
\mciteBstWouldAddEndPuncttrue
\mciteSetBstMidEndSepPunct{\mcitedefaultmidpunct}
{\mcitedefaultendpunct}{\mcitedefaultseppunct}\relax
\EndOfBibitem
\bibitem[{Krautter} \latin{et~al.}(1997){Krautter}, {Wichmann}, {Schmitt}, {Alcala}, {Neuhauser}, and {Terranegra}]{krautter1997}
{Krautter},~J.; {Wichmann},~R.; {Schmitt},~J.~H.~M.~M.; {Alcala},~J.~M.; {Neuhauser},~R.; {Terranegra},~L. {New ``weak-line''--T Tauri stars in Lupus}. \emph{\aaps} \textbf{1997}, \emph{123}, 329--352\relax
\mciteBstWouldAddEndPuncttrue
\mciteSetBstMidEndSepPunct{\mcitedefaultmidpunct}
{\mcitedefaultendpunct}{\mcitedefaultseppunct}\relax
\EndOfBibitem
\bibitem[{G{\"u}del} \latin{et~al.}(2010){G{\"u}del}, {Lahuis}, {Briggs}, {Carr}, {Glassgold}, {Henning}, {Najita}, {van Boekel}, and {van Dishoeck}]{gudel2010}
{G{\"u}del},~M.; {Lahuis},~F.; {Briggs},~K.~R.; {Carr},~J.; {Glassgold},~A.~E.; {Henning},~T.; {Najita},~J.~R.; {van Boekel},~R.; {van Dishoeck},~E.~F. {On the origin of [NeII] 12.81 {\ensuremath{\mu}}m emission from pre-main sequence stars: Disks, jets, and accretion}. \emph{\aap} \textbf{2010}, \emph{519}, A113\relax
\mciteBstWouldAddEndPuncttrue
\mciteSetBstMidEndSepPunct{\mcitedefaultmidpunct}
{\mcitedefaultendpunct}{\mcitedefaultseppunct}\relax
\EndOfBibitem
\bibitem[{Pontoppidan} \latin{et~al.}(2010){Pontoppidan}, {Salyk}, {Blake}, {Meijerink}, {Carr}, and {Najita}]{pontoppidan2010}
{Pontoppidan},~K.~M.; {Salyk},~C.; {Blake},~G.~A.; {Meijerink},~R.; {Carr},~J.~S.; {Najita},~J. {A Spitzer Survey of Mid-infrared Molecular Emission from Protoplanetary Disks. I. Detection Rates}. \emph{\apj} \textbf{2010}, \emph{720}, 887--903\relax
\mciteBstWouldAddEndPuncttrue
\mciteSetBstMidEndSepPunct{\mcitedefaultmidpunct}
{\mcitedefaultendpunct}{\mcitedefaultseppunct}\relax
\EndOfBibitem
\bibitem[{Vlasblom} \latin{et~al.}(2024){Vlasblom}, {van Dishoeck}, {Tabone}, and {Bruderer}]{vlasblom2024}
{Vlasblom},~M.; {van Dishoeck},~E.~F.; {Tabone},~B.; {Bruderer},~S. {Mid-infrared spectra of T Tauri disks: Modeling the effects of a small inner cavity on CO$_{2}$ and H$_{2}$O emission}. \emph{\aap} \textbf{2024}, \emph{682}, A91\relax
\mciteBstWouldAddEndPuncttrue
\mciteSetBstMidEndSepPunct{\mcitedefaultmidpunct}
{\mcitedefaultendpunct}{\mcitedefaultseppunct}\relax
\EndOfBibitem
\bibitem[{Vlasblom} \latin{et~al.}(2025){Vlasblom}, {Temmink}, {Sellek}, and {van Dishoeck}]{vlasblom2025_H2O}
{Vlasblom},~M.; {Temmink},~M.; {Sellek},~A.~D.; {van Dishoeck},~E.~F. {Understanding JWST water spectra: What can thermochemical models tell us about the (cold) water in protoplanetary disks?} \emph{\aap} \textbf{2025}, \emph{703}, A52\relax
\mciteBstWouldAddEndPuncttrue
\mciteSetBstMidEndSepPunct{\mcitedefaultmidpunct}
{\mcitedefaultendpunct}{\mcitedefaultseppunct}\relax
\EndOfBibitem
\bibitem[{Tabone} \latin{et~al.}(2021){Tabone}, {van Hemert}, {van Dishoeck}, and {Black}]{tabone2021}
{Tabone},~B.; {van Hemert},~M.~C.; {van Dishoeck},~E.~F.; {Black},~J.~H. {OH mid-infrared emission as a diagnostic of H$_{2}$O UV photodissociation. I. Model and application to the HH 211 shock}. \emph{\aap} \textbf{2021}, \emph{650}, A192\relax
\mciteBstWouldAddEndPuncttrue
\mciteSetBstMidEndSepPunct{\mcitedefaultmidpunct}
{\mcitedefaultendpunct}{\mcitedefaultseppunct}\relax
\EndOfBibitem
\end{mcitethebibliography}

\newpage

\section{TOC graphic}

\begin{figure}
    \centering
    \includegraphics[width=\linewidth]{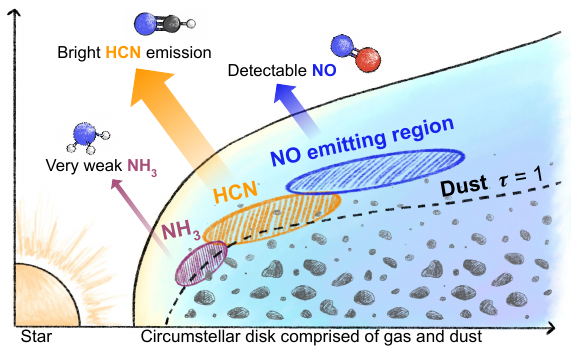}
    \caption{}
    \label{fig:toc_graphic}
\end{figure}

\end{document}